\documentclass[twocolumn,amsmath,amssymb,prl,floatfix]{revtex4-2}

\usepackage{wasysym}

\usepackage{epsfig}
\usepackage{epstopdf} 
\usepackage{pstricks}

\def\nst{four}
\def\ndt{52}
\def\npt{25}
\def\versdate{January 2026}
\def\papersversdate{December 31, 2025}

\def\al{\alpha}
\def\be{\beta}
\def\ga{\gamma}
\def\de{\delta}
\def\ep{\epsilon}
\def\ve{\varepsilon}
\def\ze{\zeta}
\def\et{\eta}
\def\th{\theta}

\def\io{\iota}
\def\ka{\kappa}
\def\la{\lambda}

\def\rh{\rho}

\def\si{\sigma}

\def\ta{\tau}
\def\up{\upsilon}
\def\ph{\phi}

\def\ch{\chi}
\def\ps{\psi}
\def\om{\omega}
\def\Ga{\Gamma}
\def\De{\Delta}

\def\La{\Lambda}
\def\Si{\Sigma}

\def\Om{\Omega}
\def\cA{{\cal A}}
\def\cB{{\cal B}}
\def\cC{{\cal C}}

\def\cK{{\cal K}}
\def\cL{{\cal L}}
\def\cO{{\cal O}}

\def\half{{\textstyle{1\over 2}}}
\def\quar{{\textstyle{1\over 4}}}
\def\eigh{{\textstyle{1\over 8}}}
\def\frac#1#2{{\textstyle{{#1}\over {#2}}}}

\def\vev#1{\langle {#1}\rangle}

\def\abs#1{\left|{#1}\right|}

\def\lsim{\mathrel{\rlap{\lower4pt\hbox{\hskip1pt$\sim$}}
    \raise1pt\hbox{$<$}}}
\def\gsim{\mathrel{\rlap{\lower4pt\hbox{\hskip1pt$\sim$}}
    \raise1pt\hbox{$>$}}}
\def\sqr#1#2{{\vcenter{\vbox{\hrule height.#2pt
         \hbox{\vrule width.#2pt height#1pt \kern#1pt
         \vrule width.#2pt}
         \hrule height.#2pt}}}}

\def\prt{\partial}
\def\Re{\hbox{Re}\,}
\def\Im{\hbox{Im}\,}
\def\re{\hbox{Re}\,}
\def\im{\hbox{Im}\,}

\def\etal{{\it et al.}}

\def\pt#1{\phantom{#1}}
\def\ol#1{\overline{#1}}

\def\kt{{\tilde\ka}}
\def\bt{{\tilde b}}
\def\ct{{\tilde c}}
\def\dt{{\tilde d}}
\def\gt{{\tilde g}}
\def\Ht{{\tilde H}}
\def\aaa{ss}
\def\bbb{\Si\Si}
\def\L{\textrm{L}}

\def\kfd#1{k_{F}^{(#1)}}
\def\kafd#1{k_{AF}^{(#1)}}
\def\klm#1#2#3{k^{(#1)}_{(#2)#3}}
\def\syjm#1#2{\phantom{}_{#1}Y_{#2}}
\def\kjm#1#2#3{k^{(#1)}_{(#2)#3}}
\def\cjm#1#2#3{c^{(#1)}_{(#2)#3}}
\def\lu#1#2#3{{(#1)_{#2}}^{#3}}

\def\kpd#1{k_{\prt}^{(#1)}}

\def\gm#1#2#3{g^{(M)}_{#1#2#3}}
\def\glA#1{g^{(A)}_{#1}}
\def\glAu#1{g^{(A)}{}^{#1}}

\def\voc{\mathrel{\rlap{\lower0pt\hbox{\hskip1pt{$c$}}}
    \raise3pt\hbox{$\neg$}}}
\def\vok{\mathrel{\rlap{\lower0pt\hbox{\hskip1pt{$k$}}}
    \raise6pt\hbox{$\neg$}}}

\def\sk#1#2#3{#1^{(#2)}_{#3}}
\def\kafoE{\sk{(\vok_{AF}^{(d)})}{1E}{njm}}
\def\kftE{\sk{(\vok_F^{(d)})}{2E}{njm}}
\def\kfoE{\sk{(\vok_F^{(d)})}{1E}{njm}}
\def\kftB{\sk{(\vok_F^{(d)})}{2B}{njm}}
\def\ring#1{{\mathaccent'27 #1}}

\def\kI{\cjm{d}{I}{jm}}
\def\kE{\kjm{d}{E}{jm}}
\def\kB{\kjm{d}{B}{jm}}
\def\kV{\kjm{d}{V}{jm}}

\def\cfdnjm#1#2{{\sk{(\voc_F^{(#1)})}{0E}{#2}}}
\newcommand{\Recfdnjm}[2]{\Re \cfdnjm{#1}{#2}}
\newcommand{\Imcfdnjm}[2]{\Im \cfdnjm{#1}{#2}}

\def\cftzE{\sk{(\voc_F^{(d)})}{0E}{njm}}
\def\kftzE{\sk{(\vok_F^{(d)})}{0E}{njm}}
\def\kftoB{\sk{(\vok_F^{(d)})}{1B}{njm}}
\def\kaftzB{\sk{(\vok_{AF}^{(d)})}{0B}{njm}}
\def\kaftoB{\sk{(\vok_{AF}^{(d)})}{1B}{njm}}

\def\gev{\mbox{ GeV}}
\def\dit#1{{\pt{---}\mbox{"}\pt{---}}}
\def\rr#1{{\cite{#1}}}
\def\da{\multicolumn{1}{c}{--}}

\def\sb{\overline{s}}
\def\mn{{\mu\nu}}
\def\lv{\checkmark}
\def\hs{@{\hspace{22pt}}}

\def\hlineone{\\[-12 pt]\hline\\[-10 pt]}
\def\hlinetwo{\\[-11 pt]\hline \\[-10 pt]}
\def\hlinethree{\\[-12 pt]\hline \\[-12 pt]}
\def\hlinesep{\\[-12 pt]\hline \\[-10 pt]}

\def\ring#1{{\mathaccent'27 #1}}
\def\cri{\ring{c}}
\def\ari{\ring{a}}

\def\sig#1{{#1}}
\def\thy#1{{#1}}

\def\cA{{\mathcal A}}
\def\cB{{\mathcal B}}
\def\cC{{\mathcal C}}

\def\widecheck#1{\hskip#1pt\huge$\check{}$}
\def\bighacek#1#2{\vbox{\ialign{##\crcr\widecheck#2\crcr
  \noalign{\kern-9.5pt\nointerlineskip}
   $\hfil\displaystyle{#1}\hfil$\crcr}}}
\def\hg{\bighacek{g}{3}}
\def\hh{\bighacek{H}{5}{}}

\def\bha{{\check b}}
\def\hc{\mbox{h.c.}}
\def\tc{\tilde{c}}
\def\tg{\tilde{g}}

\def\ol#1{\overline{#1}}

\def\prt{\partial}
\def\lrpartial{\raise 1pt\hbox{$\stackrel\leftrightarrow\partial$}}

\def\lrDmu{\stackrel{\leftrightarrow}{D_\mu}}
\def\lrDmu{{\hskip -3 pt}\stackrel{\leftrightarrow}{D_\mu}{\hskip -2pt}}

\def\lrDnu{\stackrel{\leftrightarrow}{D^\nu}}

\def\lrdmu{{\hskip -3 pt}\stackrel{\leftrightarrow}{\partial_\mu}{\hskip -2pt}}

\def\vb#1#2{e_{#1}^{{\pt{#1}}#2}}
\def\ivb#1#2{e^{#1}_{{\pt{#1}}#2}}
\def\uvb#1#2{e^{#1#2}}
\def\lvb#1#2{e_{#1#2}}

\def\nuTemplate#1#2#3#4{\big(#1^{(#2)}\big)_{#3}^{#4}}
\def\aeff#1#2#3{\nuTemplate{a_\text{eff}}{#1}{#2}{#3}}
\def\ceff#1#2#3{\nuTemplate{c_\text{eff}}{#1}{#2}{#3}}
\def\geff#1#2#3{\nuTemplate{g_\text{eff}}{#1}{#2}{#3}}
\def\Heff#1#2#3{\nuTemplate{H_\text{eff}}{#1}{#2}{#3}}

\def\aLcoef#1#2#3{\nuTemplate{a_L}{#1}{#2}{#3}}
\def\cLcoef#1#2#3{\nuTemplate{c_L}{#1}{#2}{#3}}
\def\mlcoef#1#2#3{\nuTemplate{m_l}{#1}{#2}{#3}}
\def\elcoef#1#2#3{\nuTemplate{e_l}{#1}{#2}{#3}}
\def\glcoef#1#2#3{\nuTemplate{g_l}{#1}{#2}{#3}}
\def\Hlcoef#1#2#3{\nuTemplate{H_l}{#1}{#2}{#3}}
\def\gMcoef#1#2#3{\nuTemplate{g_{M+}}{#1}{#2}{#3}}
\def\HMcoef#1#2#3{\nuTemplate{H_{M+}}{#1}{#2}{#3}}
\def\alcoef#1#2#3{\nuTemplate{a_l}{#1}{#2}{#3}}
\def\clcoef#1#2#3{\nuTemplate{c_l}{#1}{#2}{#3}}

\def\cfb#1#2{\nuTemplate{c_\text{fb}}{#1}{#2}{}}
\def\afb#1#2{\nuTemplate{a_\text{fb}}{#1}{#2}{}}
\def\gfb#1#2{\nuTemplate{g_\text{fb}}{#1}{#2}{}}
\def\cof#1#2{\nuTemplate{c_\text{of}}{#1}{#2}{}}
\def\aof#1#2{\nuTemplate{a_\text{of}}{#1}{#2}{}}
\def\cdia#1#2#3{\nuTemplate{c_\text{d}}{#1}{#2}{#3}}
\def\adia#1#2#3{\nuTemplate{a_\text{d}}{#1}{#2}{#3}}
\def\gdia#1#2#3{\nuTemplate{g_\text{d}}{#1}{#2}{#3}}
\def\cfc#1#2{\cri^{(#1)}_{#2}}
\def\afc#1#2{\ari^{(#1)}_{#2}}

\def\acfc#1#2{\ari^{(#1)}_{#2}}
\def\ccfc#1#2{\cri^{(#1)}_{#2}}

\def\template#1#2#3#4{#1^{(#2)#4}_{#3}}
\def\acoef#1#2{\template{a}{#1}{#2}{}}

\def\ccoef#1#2{\template{c}{#1}{#2}{}}

\def\gzBcoef#1#2{\template{g}{#1}{#2}{(0B)}}
\def\goBcoef#1#2{\template{g}{#1}{#2}{(1B)}}
\def\goEcoef#1#2{\template{g}{#1}{#2}{(1E)}}
\def\HzBcoef#1#2{\template{H}{#1}{#2}{(0B)}}
\def\HoBcoef#1#2{\template{H}{#1}{#2}{(1B)}}
\def\HoEcoef#1#2{\template{H}{#1}{#2}{(1E)}}

\def\nr{{\rm NR}}
\def\NR{\textrm{NR}}
\def\nrtemplate#1#2#3{#1^{\nr#3}_{#2}}
\def\anr#1{\nrtemplate{a}{#1}{}}

\def\cnr#1{\nrtemplate{c}{#1}{}}

\def\gzBnr#1{\nrtemplate{g}{#1}{(0B)}}
\def\goBnr#1{\nrtemplate{g}{#1}{(1B)}}
\def\goEnr#1{\nrtemplate{g}{#1}{(1E)}}
\def\HzBnr#1{\nrtemplate{H}{#1}{(0B)}}
\def\HoBnr#1{\nrtemplate{H}{#1}{(1B)}}
\def\HoEnr#1{\nrtemplate{H}{#1}{(1E)}}

\def\ur{{\rm UR}}
\def\urtemplate#1#2#3#4{#1^{\ur(#2)#4}_{#3}}
\def\aur#1#2{\urtemplate{a}{#1}{#2}{}}

\def\cur#1#2{\urtemplate{c}{#1}{#2}{}}

\def\gzBur#1#2{\urtemplate{g}{#1}{#2}{(0B)}}
\def\goBur#1#2{\urtemplate{g}{#1}{#2}{(1B)}}
\def\goEur#1#2{\urtemplate{g}{#1}{#2}{(1E)}}
\def\HzBur#1#2{\urtemplate{H}{#1}{#2}{(0B)}}
\def\HoBur#1#2{\urtemplate{H}{#1}{#2}{(1B)}}
\def\HoEur#1#2{\urtemplate{H}{#1}{#2}{(1E)}}

\def\ring#1{{\mathaccent'27 #1}}

\def\fctemplate#1#2#3{\ring{#1}^{(#2)}_{#3}}
\def\afc#1#2{\fctemplate{a}{#1}{#2}}

\def\cfc#1#2{\fctemplate{c}{#1}{#2}}

\def\gfc#1#2{\fctemplate{g}{#1}{#2}}
\def\Hfc#1#2{\fctemplate{H}{#1}{#2}}

\def\efc#1#2{\fctemplate{e}{#1}{#2}}
\def\ffc#1#2{\fctemplate{f}{#1}{#2}}

\def\nrfctemplate#1#2{\nrtemplate{\ring{#1}}{#2}{}}
\def\anrfc#1{\nrfctemplate{a}{#1}}

\def\cnrfc#1{\nrfctemplate{c}{#1}}

\def\gnrfc#1{\nrfctemplate{g}{#1}}
\def\Hnrfc#1{\nrfctemplate{H}{#1}}

\def\urfctemplate#1#2#3{\urtemplate{\ring{#1}}{#2}{#3}{}}
\def\aurfc#1#2{\urfctemplate{a}{#1}{#2}}

\def\curfc#1#2{\urfctemplate{c}{#1}{#2}}

\def\gurfc#1#2{\urfctemplate{g}{#1}{#2}}
\def\Hurfc#1#2{\urfctemplate{H}{#1}{#2}}

\def\ctemplate#1#2#3#4{{#1}^{(#2)#3}_{#4}}
\def\mc#1#2{\ctemplate{m}{#1}{#2}{}}
\def\mfc#1#2{\ctemplate{m}{#1}{#2}{5}}
\def\ac#1#2{\ctemplate{a}{#1}{#2}{}}
\def\bc#1#2{\ctemplate{b}{#1}{#2}{}}
\def\cc#1#2{\ctemplate{c}{#1}{#2}{}}
\def\dc#1#2{\ctemplate{d}{#1}{#2}{}}
\def\ec#1#2{\ctemplate{e}{#1}{#2}{}}
\def\fc#1#2{\ctemplate{f}{#1}{#2}{}}
\def\gc#1#2{\ctemplate{g}{#1}{#2}{}}
\def\Hc#1#2{\ctemplate{H}{#1}{#2}{}}

\def\mcf#1#2{\ctemplate{m}{#1}{#2}{F}}
\def\mfcf#1#2{\ctemplate{m}{#1}{#2}{5F}}
\def\acf#1#2{\ctemplate{a}{#1}{#2}{F}}
\def\bcf#1#2{\ctemplate{b}{#1}{#2}{F}}
\def\ccf#1#2{\ctemplate{c}{#1}{#2}{F}}
\def\dcf#1#2{\ctemplate{d}{#1}{#2}{F}}
\def\ecf#1#2{\ctemplate{e}{#1}{#2}{F}}
\def\fcf#1#2{\ctemplate{f}{#1}{#2}{F}}
\def\gcf#1#2{\ctemplate{g}{#1}{#2}{F}}
\def\Hcf#1#2{\ctemplate{H}{#1}{#2}{F}}

\def\mcpf#1#2{\ctemplate{m}{#1}{#2}{\prt F}}
\def\mfcpf#1#2{\ctemplate{m}{#1}{#2}{5\prt F}}
\def\acpf#1#2{\ctemplate{a}{#1}{#2}{\prt F}}
\def\bcpf#1#2{\ctemplate{b}{#1}{#2}{\prt F}}

\def\Hcpf#1#2{\ctemplate{H}{#1}{#2}{\prt F}}

\def\Htf#1#2#3{{\widetilde{H}}^{(#2)#3}_{\rm eff}}
\def\gtf#1#2#3{{\widetilde{g}}^{(#2)#3}_{\rm eff}}

\def\mn{{\mu\nu}}

\def\ab{{\al\be}}
\def\bec{{\be\ga}}
\def\mab{{\mu\al\be}}
\def\mnab{{\mu\nu\al\be}}
\def\abc{{\al\be\ga}}

\def\mabc{{\mu\al\be\ga}}
\def\mnabc{{\mu\nu\al\be\ga}}
\def\psb{\ol\ps{}}

\def\cpr{c^\prime}

\newcommand{\brry}{\begin{array}}
\newcommand{\erry}{\end{array}}
\def\trry#1{\brry{c}{#1}\\ \pt Q\erry}

\def\xhat{\hat X}
\def\nfrac#1#2{(#1/#2)}
\def\N#1#2{{}_{#1}{\mathcal N}_{#2}}

\def\tr{\textrm{tr}}
\def\uk{\breve{k}}
\def\LC#1#2#3{\Ga^{#1}_{{\pt{#1}}#2#3}}
\def\nsc#1#2#3{\om_{#1}^{{\pt{#1}}#2#3}}
\def\llusc#1#2#3{\om_{#1#2}^{{\pt{#1#2}}#3}}
\def\abcd#1{\al_{#1}\be_{#1}\ga_{#1}\de_{#1}}
\def\ua{\breve{a}}

\def\uc{\breve{c}}

\def\uG{\breve{G}}
\def\uH{\breve{H}}

\def\Lb{\ol{L}}
\def\Rb{\ol{R}}
\def\Qb{\ol{Q}}
\def\Ub{\ol{U}}
\def\Db{\ol{D}}

\def\TabList{1}
\def\SumMat{S2}
\def\SumPhot{S3}
\def\SumNu{S4}
\def\SumGrav{S5}

\def\DaElectrDimThreeFour{D6} 
\def\DaElectrDimFive{D7}
\def\DaElectrDimSix{D8}
\def\DaElectrDimSevenUp{D9} 
\def\DaProtonDimThree{D10} 
\def\DaProtonDimFour{D11} 
\def\DaProtonDimFive{D12} 
\def\DaProtonDimSix{D13} 
\def\DaProtonDimSevenUp{D14} 
\def\DaNeutron{D15}
\def\DaNeutronDimFive{D16} 
\def\DaNeutronDimSix{D17} 
\def\DaNeutronDimSevenUp{D18} 

\def\DaPhotDimThree{D19}
\def\DaPhotDimFour{D20}
\def\DaPhotDimFive{D21}
\def\DaPhotDimSix{D22}

\def\DaPhotDimSeven{D23}
\def\DaPhotDimEight{D24}
\def\DaPhotDimNineUp{D25}

\def\DaMatterPhoton{D26}

\def\DaMuonDimThree{D27}
\def\DaMuonDimFour{D28}
\def\DaMuonDimFiveUp{D29}
\def\DaTauDimThreeUp{D30}

\def\DaCLFCDimThree{D31}
\def\DaCLFCDimFour{D32}
\def\DaCLFCDimFive{D33}
\def\DaLQFCDimSix{D34}

\def\DaNuDimTwo{D35}
\def\DaNuDimThree{D36}
\def\DaNuDimFour{D37}
\def\DaNuDimFive{D38}
\def\DaNuDimSix{D39}

\def\DaNuDimSeven{D40}
\def\DaNuDimEight{D41}
\def\DaNuDimNineUp{D42}

\def\DaQuarkDimThree{D43}
\def\DaQuarkDimFour{D44}
\def\DaQuarkDimFiveUp{D45}

\def\DaElweak{D46}
\def\DaGluon{D47}

\def\DaGravDimTwo{D48}
\def\DaGravDimThree{D49}
\def\DaGravDimFour{D50}
\def\DaGravDimFive{D51} 
\def\DaGravDimSix{D52} 
\def\DaGravDimSevenUp{D53}

\def\DaEleGrav{D54}
\def\DaProtGrav{D55}
\def\DaNeutGrav{D56}
\def\DaMuGrav{D57}

\def\PrQEDLagr{P58}
\def\PrCPT{P59}
\def\PrFermionTilde{P60}
\def\PrInverseTilde{P61}
\def\PrPhotTilde{P62}
\def\PrSMEFermionLagr{P63}
\def\PrSMEBosonLagr{P64}
\def\PrNuCoeffs{P65}
\def\PrNuDefs{P66}
\def\PrNonMinFermiLagr{P67} 
\def\PrNonMinFermCoeffs{P68}
\def\PrNonMinPhotLagr{P69}
\def\PrNonMinPhotCoeffs{P70}
\def\PrNonMinFermPhot{P71}
\def\PrNonMinNuCoeffs{P72}

\def\PrSMEGravityGaugeLag{P73}
\def\PrSMELeptonQuarkGravityGaugeLag{P74}
\def\PrSMEHiggsGravityGaugeLag{P75}
\def\PrSMEYukawaGravityGaugeLag{P76}

\def\PrLinGrav{P77}
\def\PrLinGravSpher{P78}
\def\PrLinGravDef{P79}
\def\PrFermionGravityLag{P80}
\def\PrFermionGravityTilde{P81}
\def\PrFermionGravityNRcoeffs{P82}

\def\mbf#1{\boldsymbol #1}
\def\pd{{pd}}
\def\mr{m_{\rm r}}

\def\kb{\overline{k}{}}
\newcommand{\kei}[1]{(\kb_{\rm eff})_{#1}}

\newcommand{\ba}[1]{\left(\bar a^{#1}_\textrm{eff}\right)}
\newcommand{\bs}{\bar s}
\def\sol#1#2{{\ol s}^{(#1)}_{#2}{}} 

\def\ktrace{{\kt_{\rm tr}}}

\def\kNdjm#1#2{k^{{\rm N}(#1)}_{#2}}
\def\std#1{{\ol s}^{~(#1)}}
\def\kIdjm#1#2{\kjm{#1}{I}{#2}}
\def\kEdjm#1#2{\kjm{#1}{E}{#2}}

\def\scd{{s}^{(d)}}
\def\qbd{{q}^{(d)}}
\def\kdd{{k}^{(d)}}
\def\Sbz{{s}^{(d,1)}}
\def\Sba{{s}^{(d,2)}}
\def\Qaz{{q}^{(d,1)}}
\def\Qbz{{q}^{(d,2)}}
\def\Qaa{{q}^{(d,3)}}
\def\Qcz{{q}^{(d,4)}}
\def\Qba{{q}^{(d,5)}}
\def\Kzz{{k}^{(d,1)}}
\def\Kaz{{k}^{(d,2)}}
\def\Kbz{{k}^{(d,3)}}
\def\Kcz{{k}^{(d,4)}}

\def\qd#1#2{{q}^{(#1)}{}^{#2}}

\def\kfk{(k_{a,K}^{(5)}){}}
\def\kfdd{(k_{a,D}^{(5)}){}}
\def\kfb{(k_{a,B_d}^{(5)}){}}
\def\kfbs{(k_{a,B_s}^{(5)}){}}

\newcounter{tc1}\newcounter{tc2}
\newcounter{tr1}\newcounter{tr2}
\newlength{\h}
\def\newtableau#1#2{\psset{unit=12pt,linewidth=0.5pt}%
  \setlength{\h}{#2\psunit}\setlength{\h}{0.5\h}\addtolength{\h}{-0.3\psunit}
  \begin{pspicture}[shift=-\h](#1,#2)\small%
    \setcounter{tc1}{0}\setcounter{tc2}{1}%
    \setcounter{tr1}{#2}\setcounter{tr2}{#2}\addtocounter{tr1}{-1}%
    \psline(0,0)(0,#2)(#1,#2)}
\def\endtableau{\end{pspicture}}
\def\poxbox#1{%
  \psline(\value{tc1},\value{tr1})(\value{tc2},\value{tr1})(\value{tc2},\value{tr2})%
  \rput(\value{tc1},\value{tr1}){\rput(0.5,0.5){#1}}
  \addtocounter{tc1}{1}\addtocounter{tc2}{1}}
\def\longbox#1{%
  \addtocounter{tc2}{1}%
  \psline(\value{tc1},\value{tr1})(\value{tc2},\value{tr1})(\value{tc2},\value{tr2})%
  \rput(\value{tc1},\value{tr1}){\rput(1.0,0.5){#1}}
  \addtocounter{tc1}{2}\addtocounter{tc2}{1}}
\def\newrow{%
  \addtocounter{tr1}{-1}\addtocounter{tr2}{-1}%
  \setcounter{tc1}{0}\setcounter{tc2}{1}}

\newcommand{\beq}{\begin{equation}}
\newcommand{\eeq}{\end{equation}}
\newcommand{\bea}{\begin{eqnarray}}
\newcommand{\eea}{\end{eqnarray}}

\begin{document}

\begin{flushright}
\newpage
\vskip 0.2 truein
{\versdate}
\end{flushright}

\title{Data Tables for Lorentz and CPT Violation}

\author{V.\ Alan Kosteleck\'y$^a$
and Neil Russell$^b$}

\affiliation{$^a$Physics Department, Indiana University,
Bloomington, IN 47405\\
$^b$Physics Department, Northern Michigan University,
Marquette, MI 49855\\
\pt{xxxxxxxxxxxxxxxxxxxxxxxxxxxxxxxxxxxxxxxxxxxxxxxx}\\
\rm 
\versdate\ update of 
{\it Reviews of Modern Physics} {\bf 83}, 11 (2011)
[arXiv:0801.0287]}

\begin{abstract}
This work tabulates
measured and derived values
of coefficients for Lorentz and CPT violation
in the Standard-Model Extension.
Summary tables are extracted listing maximal attained sensitivities
in the matter, photon, neutrino, and gravity sectors.
Tables presenting definitions and properties are also compiled.
\end{abstract}

\bigskip

\maketitle

\section*{CONTENTS}

I.\ Introduction
\quad\dotfill\quad
1

II.\ Summary tables
\quad\dotfill\quad
2

III.\ Data tables
\quad\dotfill\quad
3

IV.\ Properties tables
\quad\dotfill\quad
6

\quad
A.\ Minimal QED extension
\quad\dotfill\quad
6

\quad
B.\ Minimal SME
\quad\dotfill\quad
8

\quad
C.\ Nonminimal sectors, Minkowski spacetime
\quad\dotfill\quad
8

\quad
D.\ Nonminimal sectors, Riemann spacetime
\quad\dotfill\quad
10

\quad
E.\ Linearized gravity 
\quad\dotfill\quad
11

References
\quad\dotfill\quad
13

Table \TabList:
List of tables
\quad\dotfill\quad
20

Summary tables
\SumMat--\SumGrav
\quad\dotfill\quad
23

Data tables
\DaElectrDimThreeFour--\DaMuGrav
\quad\dotfill\quad
29

Properties tables
\PrQEDLagr--\PrFermionGravityNRcoeffs
\quad\dotfill\quad
175

\section{I.\ Introduction}
\label{Introduction}

Recent years have seen a renewed interest
in experimental tests of Lorentz and CPT symmetry.
Observable signals of Lorentz and CPT violation
can be described in a model-independent way
using effective field theory \cite{kp}.

The general realistic effective field theory for Lorentz violation
is called the Standard-Model Extension
(SME) \cite{dcak,akgravity}.
It includes the Standard Model coupled to General Relativity
along with all possible operators for Lorentz violation.
Both local and global Lorentz violations are incorporated,
along with diffeomorphism, translation, and gravitational-gauge violations
\cite{kl21-1}.
Since CPT violation in realistic field theories
is accompanied by Lorentz violation \cite{owg},
the SME also describes general CPT violation.
Reviews of the SME can be found
in Refs.\ \cite{cptprocs,rb,jt,ar,qb}.

Each Lorentz-violating term
in the Lagrange density of the SME
is constructed as the coordinate-independent product
of a coefficient for Lorentz violation
with a Lorentz-violating operator.
The Lorentz-violating physics
associated with any operator is
therefore controlled by the corresponding coefficient,
and so any experimental signal for Lorentz violation can be
expressed in terms of one or more of these coefficients.
The coefficients are typically assumed 
to produce perturbative effects on known physics
in a set of concordant observer frames
\cite{kl01},
but nonperturbative effects arising from large coefficients
or in nonconcordant frames can also be considered
\cite{klss24}.

The Lorentz-violating operators in the SME
are systematically classified
according to their mass dimension,
and operators of arbitrarily large dimension can appear.
At any fixed dimension,
the operators are finite in number
and can in principle be enumerated.
A limiting case of particular interest is the minimal SME,
which can be viewed as the restriction of the SME
to include only Lorentz-violating operators of mass dimension
four or less.
The corresponding coefficients for Lorentz violation
are dimensionless or have positive mass dimension.

We compile \ndt\ {\bf\emph{data tables}} for these SME coefficients,
including both existing experimental measurements
and theory-derived limits.
Each of these data tables provides information
about the results of searches for Lorentz violation
for a specific sector of the SME.
For each measurement or constraint,
we list the relevant coefficient or combination of coefficients,
the result as presented in the literature,
the context in which the search was performed,
and the source citation.
The tables include results available from the literature
up to \papersversdate.

The scope of the searches for Lorentz violation
listed in the data tables
can be characterized roughly in terms of
depth, breadth, and refinement.
Deep searches yield great sensitivity
to a small number of SME coefficients.
Broad searches cover substantial portions
of the coefficient space,
usually at a lesser sensitivity.
Searches with high refinement
disentangle combinations of coefficients.
In the absence of a compelling signal for Lorentz violation,
all types of searches are necessary
to obtain complete coverage of the possibilities.

As a guide to the scope of the existing searches,
we extract from the data tables
\nst\ {\bf\emph{summary tables}} covering the sectors for matter
(electrons, protons, neutrons, and their antiparticles),
photons, neutrinos, and gravity.
These summary tables
list our best estimates
for the maximal attained sensitivities
to the relevant SME coefficients in the corresponding sectors.
Each entry in the summary tables
is obtained under the assumption
that only one coefficient is nonzero.
The summary tables therefore
provide information about the overall search depth and breadth,
at the cost of masking the search refinement.

In addition to the data tables and the summary tables,
we also provide \npt\ {\bf\emph{properties tables}} listing
some features and definitions
of the SME and the coefficients for Lorentz violation.
The Lagrange densities for the minimal QED extension
in Riemann spacetime,
for the minimal SME in Riemann-Cartan spacetime,
for a nonminimal Dirac fermion in Minkowski spacetime,
for the nonminimal photon sector in Minkowski spacetime,
for all gravity-gauge-lepton-quark-Higgs terms 
with operators of dimension $\leq 6$,
and for linearized gravity
are provided in tabulated form.
The mass dimensions of the operators for Lorentz violation
and their properties
under the various discrete spacetime transformations
are displayed.
Standard combinations of SME coefficients
that appear in the literature are listed.
Along with the data tables and the summary tables,
the properties tables can be used to identify
open directions for future searches.
Among these are first measurements of unconstrained coefficients,
improved sensitivities to constrained coefficients,
and studies disentangling combinations of coefficients.

The organization of the tables is as follows.
Table \TabList\ contains a list of all tables.
The \nst\ summary tables are presented next,
Tables \SumMat--\SumGrav.
These are followed by the \ndt\ data tables,
Tables \DaElectrDimThreeFour--\DaMuGrav.
The \npt\ properties tables appear last,
Tables \PrQEDLagr--\PrFermionGravityNRcoeffs.

A description of the summary tables
is given in Sec.\ II.
Information about the format and content of the data tables
is presented in Sec.\ III,
while Sec.\ IV provides an overview of
the properties tables.
The bibliography follows Sec.\ IV,
and it covers the text and all the tables.

\section{II.\ Summary tables}

The \nst\ summary tables
(Tables \SumMat--\SumGrav)
list maximal experimental sensitivities attained
for coefficients in the matter,
photon, neutrino, and gravity sectors
of the SME.
To date,
there is no confirmed experimental evidence
supporting Lorentz violation.
A few measurements suggest nonzero coefficients
at weak confidence levels.
These latter results have been excluded
in constructing the summary tables
but are listed in the data tables.
Also excluded are results based on
the reported $6\sigma$ difference
between the speeds of muon neutrinos and light
in the OPERA experiment
\cite{2011OPERA},
which has since been identified as a systematic effect
\cite{2012LVDOPERA}.

In the \nst\ summary tables,
each displayed sensitivity value represents
our conservative estimate of a $2\sigma$ limit,
given to the nearest order of magnitude,
on the modulus of the corresponding coefficient.
Our rounding convention is logarithmic:
a factor greater than or equal to $10^{0.5}$ rounds to 10,
while a factor less than $10^{0.5}$ rounds to 1.
In a few cases,
tighter results may exist when
suitable theoretical assumptions are adopted;
these results can be found in the data tables that follow.

Where observations involve a linear combination
of the coefficients appearing in the summary tables,
the displayed sensitivity for each coefficient
assumes for definiteness
that no other coefficient contributes.
Some caution is therefore advisable in applying the results
in these summary tables to situations
involving two or more nonzero coefficient values.
Care in applications is also required
because under some circumstances
certain coefficients can be intrinsically unobservable
or can be absorbed into others
by field or coordinate redefinitions,
as described in Sec.\ IV A.

\begin{figure}
\begin{center}
\centerline{
\psfig{figure=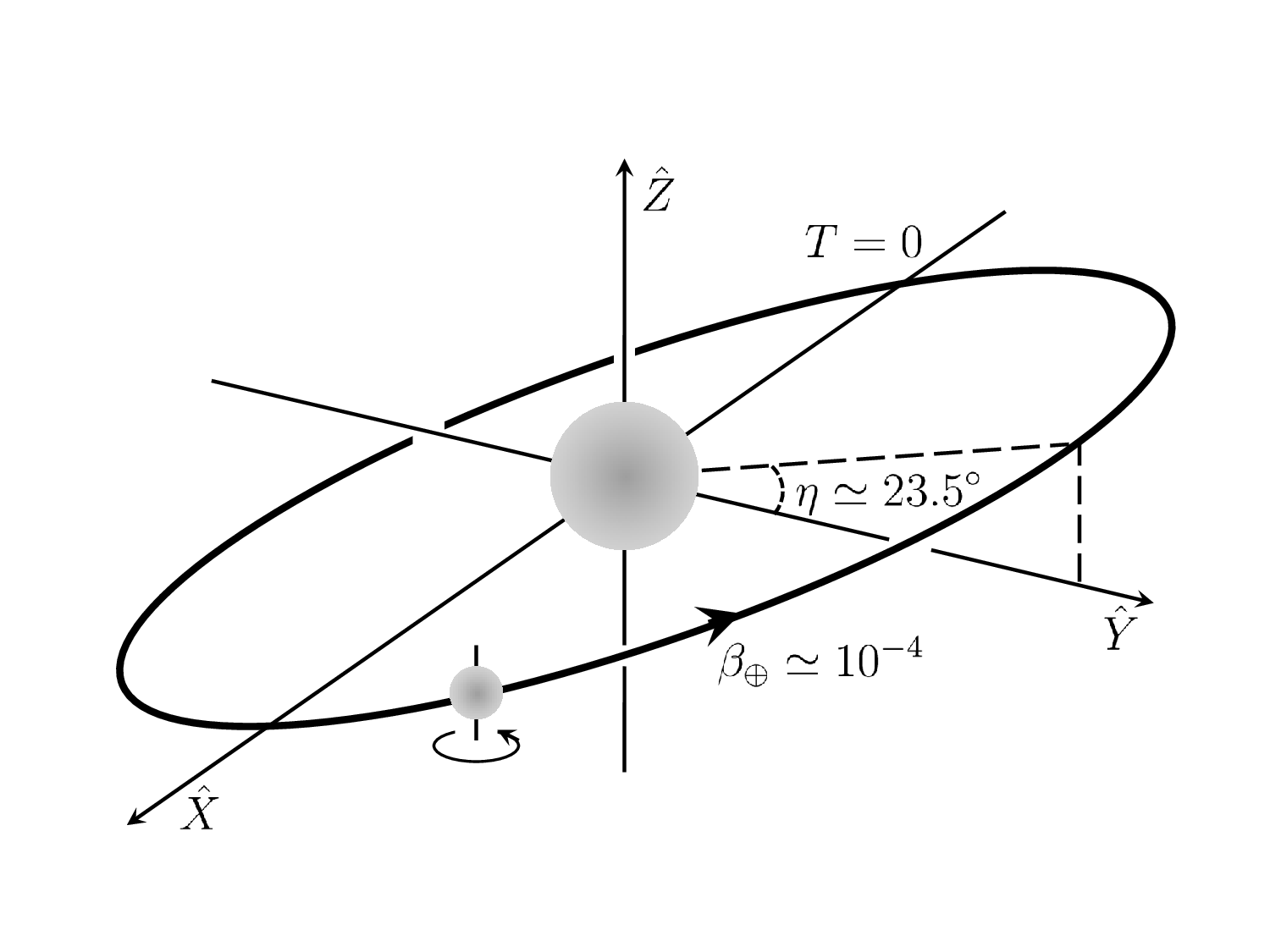,width=1.1\hsize}}
\caption{Standard Sun-centered inertial reference frame
\cite{spacepr}.}
\label{framefigure}
\end{center}
\end{figure}

In presenting the physical sensitivities,
we adopt natural units with $\hbar = c= \ep_0 = k_B = 1$
and express mass units in GeV.
Our values are reported
in the standard Sun-centered inertial reference frame
\cite{spacepr}
widely used in the literature.
This frame is illustrated in Fig.\ \ref{framefigure}.
The origin of the time coordinate $T$
is at the 2000 vernal equinox.
The $Z$ axis is directed north and parallel
to the rotational axis of the Earth at $T=0$.
The $X$ axis points from the Sun towards the vernal equinox,
while the $Y$ axis completes a right-handed system.
Some further details about this frame,
including transformations to other standard frames,
can be found in Section III A
and Appendix C of Ref.\ \cite{akmmphoton}.

Table \SumMat\ lists
the maximal attained sensitivities involving
electrons, protons, neutrons, and their antiparticles.
For each distinct massive spin-half Dirac fermion
in the minimal SME in Minkowski spacetime,
there are 44 independent observable combinations
of coefficients for Lorentz violation
in the nonrelativistic limit.
Of these, 20 also control CPT violation.
The 44 combinations are conventionally chosen
as the tilde coefficients shown.
The definitions of these 44 tilde coefficients
in terms of coefficients in the minimal SME
are listed in Table \PrFermionTilde.
All the definitions appear elsewhere in the literature
\cite{spacepr}
except the four combinations
$\tilde{b}_J^*$ and $\tilde{c}_{TT}$.
The three tilde coefficients $\tilde{b}_J^*$
are the antimatter equivalent of
the tilde coefficients $\tilde{b}_J$.
They appear in nonrelativistic studies of antimatter properties,
such as the hyperfine transitions of antihydrogen
\cite{hantih}.
The tilde coefficient $\tilde{c}_{TT}$
is a simple scaling of the coefficient $c_{TT}$
in the minimal SME,
introduced here to ensure completeness
of the set of tilde coefficients.
All tilde coefficients have dimensions of GeV in natural units.
In Table \SumMat,
a superscript indicating the particle species of relevance
is understood on all coefficients.
For example,
the first line of the table
presents limits on three different tilde coefficients,
$\tilde b_X^e$, $\tilde b_X^p$, $\tilde b_X^n$.
In the table,
a dash indicates that no sensitivity to the coefficient
has been identified to date.
A few maximal sensitivities 
are obtained by applying the inverse
of the definitions in Table \PrFermionTilde\
to data in Tables \DaElectrDimThreeFour,
\DaProtonDimThree,
and
\DaProtonDimFour.
These inverses are presented in Table \PrInverseTilde.

Table \SumPhot\ displays the maximal attained sensitivities
to coefficients for Lorentz violation
in the photon sector.
The first section of Table \SumPhot\ 
concerns those few coefficients
in the photon sector of the full SME
that control isotropic Lorentz violation
in the Sun-centered frame.
Exactly one such coefficient exists
for each operator of mass dimension $d$
\cite{2009akmmNonMin}.
The operators also violate CPT when $d$ is odd.
The table displays maximal attained sensitivities
to isotropic Lorentz violation
from operators of dimension $3\leq d \leq 9$.
The next two sections of Table \SumPhot\ 
concern the minimal photon sector.
There are 23 observable coefficient combinations for photons
in the minimal SME,
of which four also control CPT violation.
Four rows of the table
concern combinations of these four coefficients,
which have dimensions of GeV in natural units.
The remaining rows in these parts of the table
display maximal sensitivities to 10 spherical and 9 tilde coefficients,
which are convenient combinations
of the 19 dimensionless coefficients
in the minimal SME.
The definitions of all 23 combinations
of coefficients in the minimal SME
are taken from the literature
\cite{akmmphoton,2007akmm}
and are provided in Table \PrPhotTilde.
The final part of Table \SumPhot\ 
provides the 36 independent coefficients 
for the nonminimal photon sector with $d=5$,
along with existing two-sided maximal sensitivities.
The definitions of these coefficients are given in 
Refs.\ \cite{akmmphoton,2007akmm}
and are reproduced in Table \PrNonMinPhotCoeffs.

Table \SumNu\ lists the maximal attained sensitivities
to a subset of the coefficients controlling
Lorentz-violating mixing and propagation of the three flavors
$\nu_e$, $\nu_\mu$, $\nu_\ta$
of left-handed neutrinos and antineutrinos
in the SME.
In this table,
each numerical value gives the limit
on the modulus of the corresponding coefficient.
A dash indicates that no sensitivity to the coefficient
has been identified as yet.
The first 14 rows of the table concern
cartesian coefficients governing neutrino mixing
for operator dimensions $d=3$ and $d=4$.
All these flavor-mixing coefficients are complex,
so bounds on both real and imaginary parts are provided.
A subscript $e\mu$, $e\ta$, or $\mu\ta$
indicating the specific flavor mixing
is understood on each coefficient
and can be identified by the column heading.
For example,
the entry in the first row and second column
provides the constraint $|\Re(a_L)^T_{e\mu}|< 10^{-20}$ GeV.
The last eight rows of Table \SumNu\ concern spherical coefficients in the nonminimal neutrino sector,
which are defined in the literature
\cite{2011akmm}
and listed in Table \PrNonMinNuCoeffs.
These eight rows list maximal attained sensitivities
to isotropic oscillation-free operators
of dimension $3 \leq d \leq 10$
controlling propagation,
along with isotropic mixing operators of dimension $3 \leq d \leq 10$
controlling $\nu_e\leftrightarrow\nu_\mu$ 
and $\nu_\mu\leftrightarrow\nu_\ta$ oscillations.
One isotropic coefficient of each type exists
for each $d$,
and the corresponding operators violate CPT when $d$ is odd.

Table \SumGrav\ displays the maximal attained sensitivities
to certain coefficients for Lorentz violation
involving the gravity sector
of the minimal SME.
Two classes of coefficients can be distinguished
in this context:
ones appearing in the matter sector,
and ones appearing in the pure-gravity sector.
For the first class,
Table \SumGrav\ contains results for the 24 coefficients
$\ol{a}_\mu^e$, $\ol{a}_\mu^p$, $\ol{a}_\mu^n$
and $\ol{e}_\mu^e$, $\ol{e}_\mu^p$, $\ol{e}_\mu^n$
involving the electron, proton, and neutron sectors.
These observables are associated with CPT-odd operators.
The prefactor $\al$ is a model-dependent number
\cite{2008akjt}.
For the second class,
the table displays nine combinations
of the nine dimensionless coefficients
for Lorentz violation $\sb^\mn$.

\section{III.\ Data tables}

We present \ndt\ data tables compiled
from the existing literature.
Some of these
include results restricted to the minimal SME,
others list bounds on nonminimal coefficients,
and a few contain data involving both minimal and nonminimal sectors.
The \ndt\ tables cover
the electron sector
(Tables \DaElectrDimThreeFour--\DaElectrDimSevenUp),
the proton sector
(Tables \DaProtonDimThree--\DaProtonDimSevenUp),
the neutron sector
(Tables \DaNeutron--\DaNeutronDimSevenUp),
the photon sector
(Tables \DaPhotDimThree--\DaPhotDimNineUp),
nonminimal QED couplings
(Table \DaMatterPhoton),
the charged-lepton and lepton-quark sectors
(Tables \DaMuonDimThree--\DaLQFCDimSix),
the neutrino sector
(Tables \DaNuDimTwo--\DaNuDimNineUp),
the quark sector
(Tables \DaQuarkDimThree--\DaQuarkDimFiveUp),
the electroweak sector
(Table \DaElweak),
the gluon sector
(Table \DaGluon),
and the gravity sector
(Tables \DaGravDimTwo--\DaMuGrav).

Each of these \ndt\ data tables
contains four columns.
The first column lists the coefficients for Lorentz violation
or their relevant combinations.
Results for coefficients of the same generic type
are grouped together.
Certain results involve combinations of coefficients
across more than one sector;
each of these has been listed only once
in the table deemed most appropriate.
Some minor changes in notation or format
have been introduced as needed,
but for the most part the results are quoted
as they appear in the cited references.
Definitions for standard combinations of coefficients
are provided in the properties tables that follow.
A few authors use unconventional notation;
where immediate,
the match to the standard notation is shown.
Parentheses enclosing a pair of indices on a coefficient
indicate symmetrization without a factor of two.

The second column
contains the measurements and bounds,
presented in the same form as documented
in the literature.
For each generic type of coefficient,
the results are listed in reverse chronological order.
If no significant figures appear in the quoted limit
on an absolute value,
the order of magnitude of the limit is given as a power of 10.
Where both statistical and systematic errors appear
in a given result,
they are quoted in that order.

The third column contains a succinct reminder
of the physical context in which the bound is extracted,
while the fourth column contains the source citations.
The reader is referred to the latter for
details of experimental and theoretical procedures,
assumptions underlying the results,
definitions of unconventional notations,
and other relevant information.
Results deduced on theoretical grounds
are distinguished from those
obtained via direct experimental measurement
by an asterisk placed after the citation.

Tables \DaElectrDimThreeFour--\DaNeutronDimSevenUp\
contain data for the electron, proton, and neutron sectors.
These data are organized sequentially,
beginning with combinations
involving the minimal coefficients
$b_\nu$, $c_{\mn}$, $H_{\mn}$, $d_{\mn}$, $g_{\mu\nu\la}$
and followed by combinations involving nonminimal coefficients.
For all three sectors,
minimal and nonminimal coefficients are displayed in distinct tables. 
A superscript indicating the particle species of relevance
is understood on the coefficients in all these tables.
Standard definitions for the minimal coefficients
and their combinations are provided in
Tables \PrQEDLagr\ and \PrFermionTilde,
while definitions for the nonminimal coefficients are given in
Tables \PrNonMinFermiLagr\ and \PrNonMinFermCoeffs.
Many entries in Tables \DaElectrDimThreeFour--\DaNeutronDimSevenUp\
use coefficients for a single Dirac fermion written in a cartesian basis,
but some adopt instead coefficients 
decomposed using spin-weighted spherical harmonics,
which provide a convenient means of enumerating
Lorentz-violating fermion operators of arbitrary mass dimension $d$
\cite{2013AKMMfermions}. 
Some entries for the nonminimal sectors appearing in these tables
involve nonrelativistic spherical coefficients
identified by the superscript NR,
which are formed as special linear combinations 
of spherical coefficients for all dimensions $d$,
while others 
involve isotropic ultrarelativistic coefficients
denoted by a ring diacritic and the superscript UR.
One useful relation is $c_{TT}\equiv 3 \curfc{4}{}/4$.
Some results depend on $\et \simeq 23.5^{\circ}$,
which is the angle
between the equatorial and ecliptic planes in the solar system.
Note that the existing bounds on observables
involving $a_\nu^e$, $a_\nu^p$, $a_\nu^n$
and $e_\nu^e$, $e_\nu^p$, $e_\nu^n$
are obtained from gravitational experiments
and are listed with the gravity-sector results
in Table \DaGravDimThree.

Tables \DaPhotDimThree\ and \DaPhotDimFour\ present the photon-sector data
for operator dimensions $d=3$ and $d=4$,
respectively.
Most of the combinations of coefficients for Lorentz violation
appearing in the first columns of these tables are defined in
Tables \PrPhotTilde\ and \PrNonMinPhotCoeffs.
The combinations
$\klm 3 V {jm}$, $\klm 4 E {jm}$, and $\klm 4 B {jm}$
arise from analyses
\cite{2007akmm,2009akmmNonMin,2008akmmAstroph}
using spin-weighted spherical harmonics.
The factor of $\be_\oplus$ appearing in some places
is the speed of the Earth
in the standard Sun-centered reference frame,
which is about $10^{-4}$ in natural units.

Tables \DaPhotDimFive--\DaPhotDimNineUp\
contain a compilation of some measurements and bounds
on coefficients for Lorentz violation
in the nonminimal photon sector of the SME.
Results are available for a variety of operators
of dimensions greater than 4,
with the coefficients for
$d=5$, $d=6$, $d=7$, $d=8$, and $d\geq 9$
being placed in distinct tables.
The spin-weighted spherical harmonics
also provide a convenient basis for classifying photon operators
of arbitrary mass dimension $d$
\cite{2009akmmNonMin}.
The corresponding coefficients in the photon sector
are listed in
Table \PrNonMinPhotCoeffs.
Some constraints have been obtained for
the vacuum coefficients for Lorentz violation,
which are
$\cjm d I {jm}$, $\klm d E {jm}$, $\klm d B {jm}$
for even $d$
and $\klm d V {jm}$ for odd $d$,
where the subscripts $jm$
label the angular-momentum quantum numbers.
In the first columns of
Tables \DaPhotDimFive--\DaPhotDimNineUp,
the various spherical harmonics $_sY_{jm}(\th,\ph)$ of spin weight $s$
are evaluated at specified angles,
which are the celestial coordinates
of certain astrophysical sources.
The conventional spherical harmonics of spin weight 0
are denoted as $Y_{jm}(\th,\ph)$.
Measurements are also listed
of some vacuum-orthogonal coefficients for Lorentz violation,
which are distinguished by a negation diacritic $\neg$.

Table \DaMatterPhoton\ 
provides constraints on nonminimal QED interactions 
in the SME Lagrange density.
Some of these couple the electromagnetic field strength $F_{\mu\nu}$ 
to fermion bilinear operators,
and the corresponding coefficients
are distinguished by a subscript $F$.
Others involve 4-fermion interactions.
The possible terms for $d=5$ and 6 
are given in Table \PrNonMinFermPhot.
Constraints have been obtained for several different particle species,
and many coefficients in Table \DaMatterPhoton\ 
are accordingly labeled with a subscript denoting the flavor.
Some of the entries in the first column 
involve bounds on special limits of these coefficients,
for which the definitions are given in the corresponding references. 

Tables \DaMuonDimThree--\DaCLFCDimFive\
list measurements and bounds
on coefficients for Lorentz violation
involving the second- and third-generation charged leptons.
Results for the muon and its antiparticle
for operator dimensions $d=3$, $d=4$, and $d\geq 5$
are collected in Tables \DaMuonDimThree,
\DaMuonDimFour, and \DaMuonDimFiveUp,
respectively.
These results are expressed either using the cartesian coefficients 
specified in the lepton sector of Table \PrSMEFermionLagr\
or via the spherical basis for Dirac fermions
\cite{2013AKMMfermions}
using the spherical coefficients summarized in Table \PrNonMinFermCoeffs.
Some nonrelativistic experiments are sensitive 
to special linear combinations of spherical coefficients 
involving all dimensions $d$,
which are identified by the superscript NR
and have properties listed in Table \PrNonMinFermCoeffs.
The results for these nonrelativistic coefficients
are placed in the tables according to their mass dimension
and are separated from other coefficients of definite $d$
by a horizontal line. 
A few rows of the muon data tables present limits on nonminimal operators
governed by isotropic ultrarelativistic coefficients,
denoted by a ring diacritic and the superscript UR,
which are also listed in Table \PrNonMinFermCoeffs.
Experiments with storage rings studying the muon anomalous magnetic moment
are sensitive to combinations of spherical coefficients at each $d$
that involve boost factors
and are denoted with a h\'a\v cek diacritic.
These h\'a\v cek coefficients are defined in Table XI 
of Ref.\ \cite{2014gkvMuons}.
The few existing results for the tau and its antiparticle
can be found in Table \DaTauDimThreeUp.
Flavor-changing operators also appear in the charged-lepton 
and the lepton-quark sectors.
Constraints on coefficients governing the corresponding effects 
are listed in Tables \DaCLFCDimThree--\DaLQFCDimSix.

Tables \DaNuDimTwo--\DaNuDimNineUp\
collect results involving coefficients in the neutrino sector.
Limits for neutrino operators of renormalizable dimension
appear in Tables \DaNuDimTwo--\DaNuDimFour, 
while Tables \DaNuDimFive--\DaNuDimNineUp\
contain results for coefficients in the nonminimal neutrino sector.
Most results are for operators of dimension $2\leq d \leq 10$,
but a few results in Table \DaNuDimNineUp\
concern operators of arbitrary dimension.
In compiling the results,
a separate table is used for each value of $d$ in the range
$2\leq d \leq 8$ and for $d\geq 9$.
Many of the cartesian coefficients appearing 
in the first columns of these tables
are provided in Table \PrNuDefs.
For a general treatment,
in parallel with the fermion and photon sectors,
a basis using spin-weighted spherical harmonics is useful 
to classify the neutrino operators
\cite{2011akmm}.
The corresponding sets of spherical coefficients are
listed in Table \PrNonMinNuCoeffs.
In Tables \DaNuDimTwo--\DaNuDimNineUp,
oscillation-free coefficients are listed 
before those controlling neutrino mixing,
with a horizontal line clarifying the separation where useful. 
In Table \DaNuDimNineUp,
horizontal lines also separate
coefficients of different $d$.
The final section of Table \DaNuDimNineUp\
concerns more general models
and presents constraints involving combinations
of an infinite number of coefficients.
In some entries in these tables,
orientation information is encoded in the factors
${}_{s}{\mathcal N}_{jm}$,
which are defined in Eq.\ (93) of Ref.\ \cite{2011akmm}
and evaluated at specific angles
determined by the neutrino propagation direction
and the laboratory location.
In all these tables,
many of the results listed
are obtained in the context of specific neutrino models,
as described in the cited papers.

Tables \DaQuarkDimThree--\DaQuarkDimFiveUp\ present
experimental sensitivities to coefficients for operators
in the quark sector.
Some coefficients appearing in these tables
are obtained from experiments with mesons
and are composite quantities defined in the corresponding references.
They are effective coefficients
for which complete analytical expressions are as yet unknown,
formed from certain quark-sector coefficients
appearing in Table \PrSMEFermionLagr\
and from other quantities arising
from the quark binding in the mesons.
A few rows of Table \DaQuarkDimFour\ contain bounds 
on isotropic ultrarelativistic spherical coefficients
\cite{2013AKMMfermions}, 
denoted by a ring diacritic and the superscript UR.
Properties of these coefficients are provided 
in Table \PrNonMinFermCoeffs.

Tables \DaElweak\ and \DaGluon\
concern coefficients in the gauge sectors
of the minimal SME.
Results for the electroweak sector are listed
in Table \DaElweak,
while those for the gluon sector are in Table \DaGluon.
The coefficients for the electroweak sector
are defined in the gauge and Higgs sections of
Table \PrSMEBosonLagr.
Each gluon-sector coefficient
is the analogue of the corresponding
photon-sector coefficient
defined in Table \PrPhotTilde.

Tables \DaGravDimTwo--\DaMuGrav\
present measurements and bounds concerning
operators in the minimal and nonmininmal gravity sector of the SME.
Tables \DaGravDimTwo\ and \DaGravDimThree\
present results on coefficients for diffeomorphism and Lorentz violation
in the linearized pure-gravity sector
\cite{2017km}.
Table \DaGravDimThree\ 
also includes limits on the matter-sector coefficients 
$a_\nu^e$, $a_\nu^p$, $a_\nu^n$,
and $e_\nu^e$, $e_\nu^p$, $e_\nu^n$,
which arise at $d=3$ and $d=4$
and are detectable in gravitational experiments.
Table \DaGravDimFour\
involves coefficients in the pure-gravity sector
of the minimal SME,
all of which occur at $d=4$.
The specific combinations 
appearing in some entries in the first column
are defined in the references
and are expressed in terms of the coefficients
for Lorentz violation listed
in the gravity section of Table \PrSMEBosonLagr.
Tables \DaGravDimFive--\DaGravDimSevenUp\ 
list existing experimental constraints
on coefficients in the nonminimal pure-gravity sector.
They include measurements of
combinations of coefficients with $d=5$ and $d=6$
that are observable in nonrelativistic 
and post-Newton experiments.
These coefficients are defined in Eq.\ (7) of Ref.\ \cite{2014bkx}
and Eq.\ (3) of Ref.\ \cite{2016kmgw}. 
Tables \DaGravDimFive--\DaGravDimSevenUp\ 
also display constraints on 
effective spherical coefficients with values $d\leq 10$
that govern Lorentz violation affecting 
the propagation of gravitational waves.
These effective spherical coefficients are defined 
in Eq.\ (7) of Ref.\ \cite{2016kmgw} 
and Eq.\ (45) of Ref.\ \cite{2015akjtCerenkov}.

Tables \DaEleGrav--\DaMuGrav\
list constraints on coefficients governing the fermion-gravity sector
in the limit where gravity is linearized about Minkowski spacetime.
Relevant terms of mass dimension $d\leq 5$ 
in the corresponding Lagrange density
are provided in the properties Table \PrFermionGravityLag.
Each of the four tables \DaEleGrav--\DaMuGrav\
is devoted to a single fermion flavor,
including electrons, protons, neutrons, and muons.
Many of the bounds listed are deduced
from comparisons of experimental results at different elevations,
while others are obtained from laboratory measurements.
Some of the tilde combinations displayed in these tables 
contain a repeated index $\Si\Si$,
which denotes summation over the individual tilde combinations 
with indices $TT$, $XX$, $YY$, $ZZ$ in the Sun-centered frame.
The relationships between these tilde combinations
and terms in the Lagrange density for linearized gravity
are given in the properties Table \PrFermionGravityTilde.
Some of the constraints listed in Tables \DaEleGrav--\DaMuGrav\
are expressed instead in terms of the nonrelativistic coefficients
that appear in the nonrelativistic hamiltonian
describing the coupling of fermions to the metric fluctuation
in linearized gravity.
These nonrelativistic coefficients are displayed 
in the properties Table \PrFermionGravityNRcoeffs.

\section{IV.\ Properties tables}

The \npt\ properties tables list various features and definitions
related to Lorentz violation.
Five tables concern
the terms in the restriction of the minimal SME
to quantum electrodynamics (QED) in Riemann spacetime.
For this theory,
which is called the minimal QED extension,
the tables include information about
the operator structure
(Table \PrQEDLagr),
the action of discrete symmetries
(Table \PrCPT),
and some useful coefficient combinations
(Tables \PrFermionTilde--\PrPhotTilde).
Two tables contain information about
the matter sector (Table \PrSMEFermionLagr)
and the gauge and gravity sectors
(Table \PrSMEBosonLagr)
of the minimal SME in Riemann-Cartan spacetime.
Another two tables
summarize some features of the coefficients for Lorentz violation
concerning operators of renormalizable dimension
in the neutrino sector.
One (Table \PrNuCoeffs)
lists the cartesian coefficients and some of their properties,
while the other (Table \PrNuDefs)
provides the connection
between cartesian and spherical coefficients.
Six tables
(Tables \PrNonMinFermiLagr--\PrNonMinNuCoeffs)
display information about the operator structure
and the spherical coefficients for Lorentz violation
in the nonminimal fermion, photon, and neutrino sectors
in Minkowski spacetime.
Four tables
(Tables \PrSMEGravityGaugeLag--\PrSMEYukawaGravityGaugeLag)
provide all terms with mass dimension $\leq 6$
in the Lagrange density of the SME in Riemann spacetime. 
The remaining tables
(Tables	\PrLinGrav--\PrFermionGravityNRcoeffs)
concern the gravity sector in the linearized limit,
including pure-gravity terms and fermion-gravity couplings.

For these properties tables,
our primary conventions are those of Ref.\ \cite{akgravity}.
Greek indices $\mu, \nu, \la, \ldots$
refer to curved-spacetime coordinates
and Latin indices $a, b, c, \ldots$
to local Lorentz coordinates.
The vierbein formalism
\cite{vierbein},
which relates the two sets of coordinates,
is adopted to facilitate the description of spinors
on the spacetime manifold.
The determinant $e$ of the vierbein $\vb\mu a$ is related
to the determinant $g$ of the metric $g_{\mu\nu}$
by $e = \sqrt{-g}$.
The conventions for the Dirac matrices $\ga^a$
are given in Appendix A of Ref.\ \cite{akgravity}.
The Newton gravitational constant $G_N$
enters as the combination $\ka \equiv 8 \pi G_N$,
and it has dimensions of inverse mass squared.

In the Minkowski-spacetime limit,
the metric $g_{\mu\nu}$ is written $\et_{\mu\nu}$
with diagonal entries $(-1, 1,1,1)$.
For decompositions into time and space components,
we adopt the Sun-centered frame
of Fig.\ \ref{framefigure}
and use indices $J, K, L, \ldots$ to
denote the three spatial components $X, Y, Z$.
The sign of the antisymmetric tensor $\ep_{\ka\la\mu\nu}$
is fixed via the component $\ep_{TXYZ} = +1$,
and the antisymmetric symbol in three spatial dimensions
is defined with $\ep_{XYZ} = +1$.
Note that some of the literature
on the SME in Minkowski spacetime
adopts a metric $\et_{\mu\nu}$ of opposite sign,
following the common present usage in quantum physics
instead of the one in relativity.
Under this alternative convention,
terms in the Lagrange density
with an odd number of index contractions
have opposite signs to those appearing in this work.
The numerical results for the SME coefficients in the tables
are unaffected by the convention.

\subsection{A.\ \ Minimal QED extension}

Table \PrQEDLagr\ concerns
the minimal QED extension,
for which the basic nongravitational fields
are a Dirac fermion $\ps$ and the photon $A_\mu$.
The electromagnetic field-strength tensor is
$F_{\mu\nu} = \prt_\mu A_\nu-\prt_\nu A_\mu$.
The pure-gravity sector involves
the Riemann tensor $R_{\ka\la\mu\nu}$,
the Ricci tensor $R_{\mu\nu}$,
the curvature scalar $R$,
and the cosmological constant $\La$.
The spacetime covariant derivative $D_\mu$
corrects local Lorentz indices using the spin connection,
corrects spacetime indices using the Cartan connection,
and contains the usual gauge field $A_\mu$ for the photon.
The notation $\lrDmu$
is an abbreviation for the difference of two terms,
the first with derivative acting to the right
and the second acting to the left.
Note that Table \PrQEDLagr\ is restricted to the zero-torsion limit of the minimal SME.
The general case \cite{akgravity}
involves additional operators constructed
with the torsion tensor $T^\al_{\pt{\al}\mu\nu}$.
The Minkowski-spacetime limit of QED with nonzero torsion
contains terms that mimic Lorentz violation,
so searches for Lorentz violation can be used
to bound components of the torsion tensor
\cite{2008krt,2013lys}.
A similar line of reasoning leads to constraints
on components of the nonmetricity tensor
from bounds on Lorentz violation
\cite{2016Foster,2017Lehnert}.

In Table \PrQEDLagr,
each line specifies one term
in the Lagrange density for the QED extension
in Riemann spacetime.
Both conventional QED terms
and ones with Lorentz violation are included.
The first column indicates the sector
to which the term belongs.
The second column lists the coefficient controlling
the corresponding operator.
Note the standard use of an upper-case letter
for the coefficient $H_{\mu\nu}$,
which distinguishes it from
the metric fluctuation $h_{\mu\nu}$.
The third column shows the number of components
for the coefficient.
The next three columns list the operator,
its mass dimension,
and the vierbein factor contracting the coefficient and the operator.
The final two columns list the properties
of the term under CPT and Lorentz transformations.
A CPT-even operator is indicated by a plus sign
and a CPT-odd one by a minus sign,
while terms violating Lorentz invariance
are identified by a check mark.

As an example,
consider the fourth row of Table \PrQEDLagr.
This concerns the term in the fermion sector
with coefficient $a_\mu$ for Lorentz violation.
The coefficient has four independent components,
which control the four Lorentz-violating operators
$\overline\ps \ga^a \ps$.
The gravitational couplings of this operator
are contained in the vierbein product $e e^\mu{}_a$.
The corresponding term in the Lagrange density
for the minimal QED extension in Riemann spacetime
is $\cL_a = - e a_\mu e^\mu{}_a \overline\ps \ga^a\ps$.
It has mass dimension 3 and is CPT odd.
The Minkowski-spacetime limit of this term
can be obtained by the vierbein replacement
$e_\mu{}^a \to \de_\mu{}^a$.
The number of index contractions in $\cL_a$ is two,
one each for the $\mu$ and $a$ indices,
so the overall sign of $\cL_a$ is unaffected
by the choice of convention for the Minkowski metric.

The properties listed in Table \PrQEDLagr\ are those of the operators in the Lagrange density
rather than those associated with observables.
The issue of observability of a given coefficient can be subtle
because experiments
always involve comparisons of at least two quantities.
The point is that in certain tests
a given coefficient may produce the same effect
on two or more quantities and so may be unobservable,
or it may produce effects indistinguishable
from those of other coefficients.
This situation can often be theoretically understood
via a field redefinition that eliminates the coefficient
from the relevant part of the Lagrange density
without affecting the dynamics of the experiment in question.
For example,
a constant coefficient $a_\mu$
in the minimal QED extension in Minkowski spacetime
is unobservable in any experiment involving a single fermion flavor
because it can be absorbed as a phase shift in the fermion field
\cite{dcak}.
The situation changes in Riemann spacetime,
where three of the four components of $a_\mu$ become observables
affecting the gravitational properties of the fermion
\cite{akgravity}.
Another example is provided by the coefficient $f_\mu$
in the minimal QED extension in Minkowski spacetime,
which can be converted into a coefficient
of the $c_{\mu\nu}$ type via a change of spinor basis
\cite{2006Altschul-f,2010aknr}.
Additional subtleties arise because
any experiment must always choose
definitions of clock ticking rates, clock synchronizations,
rod lengths, and rod isotropies.
This involves 10 free coordinate choices
and implies the unobservability of 10 combinations
of coefficients for Lorentz violation
\cite{2010akjt}.

Table \PrCPT\ lists the properties
under discrete-symmetry transformations
of the Lorentz-violating operators in the minimal QED extension
\cite{2002klp}.
The seven transformations considered are
charge conjugation C,
parity inversion P,
time reversal T,
and their combinations
CP, CT, PT, and CPT.
The first column specifies the operator
by indicating its corresponding coefficient.
Each of the other columns
concerns one of the seven transformations.
An even operator is indicated by a plus sign
and an odd one by a minus sign.
The table contains eight rows,
one for each of the eight possible combinations
of signs under C, P, and T.

Tables \PrFermionTilde\ and \PrInverseTilde\ concern
the 44 combinations of coefficients for Lorentz violation
that frequently appear in experimental analyses
involving the fermion sector of the minimal QED extension
in Minkowski spacetime
in the nonrelativistic limit.
These combinations are conventionally denoted by tilde coefficients,
and they are defined by the first two columns of
Table \PrFermionTilde.
Note that six of these combinations,
$\tc_X$, $\tc_Y$, $\tc_Z$,
$\tg_{TX}$, $\tg_{TY}$, and $\tg_{TZ}$,
are denoted as
$\tc_{Q,Y}$, $\tc_{Q,X}$, $\tc_{XY}$,
$\tg_{Q,Y}$, $\tg_{Q,X}$, and $\tg_{XY}$,
respectively,
in some early publications.
The definitions in the table
are given for a generic fermion of mass $m$.
Most applications in the literature involve
electrons, protons, neutrons, and their antiparticles,
for which the corresponding mass is understood.
The final column lists the number of independent components
of each coefficient.
For matter involving
electrons, protons, neutrons, and their antiparticles,
there are therefore 132 independent
observable coefficients for Lorentz violation
in the minimal QED sector of the SME
in Minkowski spacetime.

The 44 tilde coefficients in Table \PrFermionTilde\
are formed from 44 linear combinations
of the 76 available coefficients
in the minimal SME
restricted to a single free fermion in flat spacetime.
In this limit,
the other 32 degrees of freedom can be removed
by field redefinitions.
Table \PrInverseTilde\
lists each of these 44 linear combinations
and gives its equivalent in terms of tilde coefficients.
For brevity,
some of the 44 linear combinations are expressed
using the definitions
\bea
d^\pm_{\mu\nu} &=& \frac 1 2 (d_{\mu\nu}\pm d_{\nu\mu}),
\quad
\glAu\mu =
\frac 1 6 \ep^{\al\be\ga\mu} g_{\al\be\ga},
\nonumber\\
\gm \mu\nu\la &=&
\frac 1 3 (g_{\mu\nu\la} + g_{\mu\la\nu}
+ \et_{\mu\la} g_{\nu\al\be} \et^{\al\be})
- (\mu \leftrightarrow \nu).
\nonumber
\eea
The map between the two coefficient sets
can be block-diagonalized
\cite{2012Fittante},
and this is reflected in Table \PrInverseTilde\
by horizontal lines separating each block.
The number of independent degrees of freedom
in each block is provided in the third column of the table.

Table \PrPhotTilde\ presents definitions for certain combinations
of the 23 coefficients for Lorentz violation
in the photon sector of the minimal QED extension
in Minkowski spacetime.
This table has five sections.
The first section consists of five rows
listing 19 widely used combinations
of the 19 coefficients for CPT-even Lorentz violation.
The second section provides 10 alternative combinations
involving the 10 CPT-even Lorentz-violating operators
relevant to leading-order birefringence
\cite{akmmphoton}.
The remaining three sections list four combinations
of the four coefficients for CPT-odd Lorentz violation,
followed by 19 combinations
of the 19 coefficients for CPT-even Lorentz violation.
The 23 combinations in these last three sections
appear when a basis of spin-weighted spherical harmonics is adopted.

\subsection{B.\ \ Minimal SME}

Table \PrSMEFermionLagr\ concerns the fermion-sector terms
in the Lagrange density
of the minimal SME in Riemann-Cartan spacetime.
The column headings are similar to those
in Table \PrQEDLagr.
In the lepton sector,
the left- and right-handed leptons
are denoted by $L_A$ and $R_A$,
where $A$ is the generation index.
The $SU(2)$ doublet $L_A$
includes the three neutrino fields
$\nu_e$, $\nu_\mu$,  $\nu_\ta$
and the left-handed components of the three charged leptons
$e$, $\mu$, and $\ta$.
The $SU(2)$ singlet $R_A$
contains the right-handed components of
$e$, $\mu$, and $\ta$.
The derivative $D_\mu$
is both spacetime
and $SU(3)\times SU(2)\times U(1)$ covariant.
The quark fields are denoted $U_A$, $D_A$, and $Q_A$,
where $A$ is the generation index.
The right-handed components of the $u$, $c$, and $t$ quarks
are the $SU(2)$ singlets $U_A$,
while the right-handed components of $d$, $s$, and $b$
are the $SU(2)$ singlets $D_A$.
The six left-handed quark fields
are contained in the $SU(2)$ doublet $Q_A$.
The Yukawa sector involves terms coupling the Higgs doublet $\ph$
to the leptons and to the quarks.
The conventional Yukawa-coupling matrices
are denoted $(G_L)_{AB}$, $(G_U)_{AB}$, and $(G_D)_{AB}$.
The hermitian conjugate of an operator
is abbreviated h.c.\ in the table.

Table \PrSMEBosonLagr\ presents information about
the Higgs, gauge, and pure-gravity sectors
for the Lagrange density
of the minimal SME in Riemann-Cartan spacetime.
The structure of the table is the same
as that of Table \PrSMEFermionLagr.
As before,
$D_\mu$ is both a spacetime
and an $SU(3)\times SU(2)\times U(1)$ covariant derivative.
The complex Higgs field is denoted $\ph$,
the $SU(3)$ color gauge fields
and the $SU(2)$ gauge fields
are the hermitian adjoint matrices
$G_\mu$ and $W_\mu$,
respectively,
while the $U(1)$ hypercharge gauge field
is the singlet $B_\mu$.
The coupling constants for SU(3), SU(2), and U(1)
are $g_3$, $g$, and $g^\prime$, respectively.
The charge $q$ for the electromagnetic U(1) group
is given in terms of $g$, $g^\prime$,
and the angle $\th_W$ by 
$q = g \sin \th_W = g^\prime \cos \th_W$.
Each gauge field has an associated field strength,
denoted $G_{\mu\nu}$ for the strong interactions,
$W_{\mu\nu}$ for the weak interactions,
and
$B_{\mu\nu}$ for the hypercharge.
The pure-gravity sector of Table \PrSMEBosonLagr\ differs from that in Table \PrQEDLagr\ only in the addition of terms
involving the torsion field $T^\al_{\pt{\al}\mu\nu}$.

The minimal SME in Riemann-Cartan spacetime
described in Tables \PrSMEFermionLagr\ and \PrSMEBosonLagr\ can be reduced to
the minimal QED in Riemann spacetime
described in Table \PrQEDLagr\ as follows.
For the gauge sector,
including the covariant derivatives,
remove all the gauge fields except
the charge $U(1)$ field
in the photon limit $B_\mu \rightarrow A_\mu$,
and remove all the Higgs terms.
For the gravity sector,
remove all the torsion terms.
For the fermion sector,
restrict the lepton generation index to a single value,
remove all quark and neutrino terms,
and replace the Yukawa-coupling terms
with the relevant fermion mass terms.

Table \PrNuCoeffs\ concerns
the piece of the neutrino sector in the SME
that contains operators of renormalizable dimension,
including both neutrino masses and Lorentz-violating terms.
We restrict attention to three generations
of active neutrinos and antineutrinos,
allowing for possible violations
of $SU(3)\times SU(2)\times U(1)$ gauge symmetry
and lepton number
\cite{2004akmmnu}.
In the table,
the first row involves the usual neutrino mass matrix
$(m_l m_l^\dagger)_{ab}$,
where the indices $a, b$ take values
$e$, $\mu$, and $\ta$.
The next four rows concern
coefficients for Lorentz violation
that enter without mass in the relativistic limit
of the effective hamiltonian for neutrino propagation and mixing,
while the final five rows
list coefficients that involve
combinations of mass and Lorentz-violating effects
\cite{2011akmm}.
The first column labels the coefficient types,
the second column lists the coefficients,
and the third gives the dimensions
of the corresponding operators.
The fourth column indicates generically
the type of neutrino oscillations controlled by the coefficients.
The final two columns list the properties
of the operators under CPT and Lorentz transformations.

Table \PrNuDefs\ provides the relationships between
the spherical coefficients
for neutrino operators of renormalizable dimension
listed in Table \PrNonMinNuCoeffs\ 
and the cartesian ones listed in Table \PrNuCoeffs.
The spherical coefficients for Lorentz violation
arise when a basis of spin-weighted spherical harmonics is chosen
\cite{2011akmm}.
Table \PrNuDefs\ is split into six sections according to the
dimension and CPT properties of the operators,
with each section having three columns.
The first column lists the spherical coefficients,
the second shows the corresponding combination
of cartesian components,
and the third shows the number of independent
spherical coefficients appearing in each section.
On each spherical coefficient,
superscript flavor-space indices $ab$ are understood.
To keep the expressions compact,
some of the combinations are expressed
using cartesian components of the complex vector
$\hat{X}_\pm = \hat{X} \mp i \hat{Y}$.
The counting of independent spherical coefficients
is obtained by taking into account hermiticity,
symmetry, or antisymmetry conditions in flavor space.
In total,
369 independent spherical coefficients exist
for neutrino operators of renormalizable dimension.

\subsection{C.\ \ Nonminimal sectors, Minkowski spacetime}

Tables \PrNonMinFermiLagr\ and \PrNonMinFermCoeffs\
present information about the nonminimal quadratic sector 
for a single Dirac fermion in Minkowski spacetime
\cite{2013AKMMfermions}.
Contributions to the Lagrange density at arbitrary $d$ are provided in
Table \PrNonMinFermiLagr.
The structure of this table is similar to that
adopted for Tables \PrQEDLagr,
\PrSMEFermionLagr, and \PrSMEBosonLagr,
with each row associated with a term in the Lagrange density.
The entries in the first two columns 
display the coefficients for Lorentz violation
in the cartesian basis
and the number of independent degrees of freedom they contain. 
The third column shows the hermitian operator
to which each coefficient corresponds,
while the fourth gives the operator mass dimension $d$.
The conventional multiplicative factor
for the coefficient and the operator 
is given in the next column.
The final two columns specify the CPT handedness
and the Lorentz property of the operator,
using the same conventions as Table \PrQEDLagr.

Table \PrNonMinFermiLagr\ is split into two pieces 
by a horizontal line.
The entries above the line represent conventional
Lorentz-invariant contributions of renormalizable dimension
to the Lagrange density.
The row containing $m_5$ is included for generality,
although in the absence of a chiral anomaly the corresponding term
in the Lagrange density can be absorbed into the usual mass $m$
via a chiral transformation.
The entries below the horizontal line 
concern operators of arbitrary mass dimensions $d$.
Most of these operators are Lorentz violating,
but a few trace components are Lorentz invariant.
The value of $d$ for each coefficient is indicated by a superscript, 
and the coefficient itself has mass dimension $4-d$.
For $d=3$ and $d=4$,
the entries reduce to those of the minimal theory
given in Table \PrQEDLagr.
Note that none of the operators associated with the coefficients 
$m^{(d)\al_1\ldots\al_{d-3}}$ 
and $m_5{}^{(d)\al_1\ldots\al_{d-3}}$ 
are of renormalizable dimension.

Table \PrNonMinFermCoeffs\ contains properties 
of the spherical coefficients for Lorentz violation
in the nonminimal quadratic sector
for a single Dirac fermion in Minkowski spacetime.
The spherical coefficients have 
comparatively simple transformation properties under rotations. 
The table has six sections separated by horizontal lines,
each representing a specific theoretical scenario
labeled in the first column.
The second and third columns contain the spherical coefficients
and the dimensions $d$ at which they occur.
The next two columns show the allowed ranges
of the subscripts on the coefficients,
while the final column lists the number
of independent components appearing in each coefficient.

Allowing for operators of arbitrary mass dimension,
the general effective hamiltonian
describing the propagation of a Dirac fermion
is formed from the eight sets of spherical coefficients  
that are listed in the first section of
Table \PrNonMinFermCoeffs.
The subset of these coefficients 
that is isotropic in a fixed inertial frame
is given in the second section of this table.
The superscript $d$ denotes the mass dimension of the corresponding operator,
while the subscripts $n$, $j$, $m$
specify the energy dependence,
the total angular momentum,
and the azimuthal component of the angular momentum,
respectively.
The superscripts $E$ and $B$ fix the parity of the corresponding operator,
while the numerals 0, 1, or 2
preceding $E$ or $B$ provide its spin weight.
For a nonrelativistic Dirac fermion,
only certain linear combinations of coefficients
appear in the effective hamiltonian, 
and the physics is governed by  
the eight smaller sets of nonrelativistic coefficients
listed in the third section of the table.
The isotropic restriction of these coefficients 
is shown in the fourth section.
Similarly,
only a subset of the general spherical coefficients
appears in the effective hamiltonian for a high-energy fermion,
and these ultrarelativistic coefficients
and their isotropic restriction are displayed
in the final two sections of the table.

Table \PrNonMinPhotLagr\ and \PrNonMinPhotCoeffs\ 
provide information about the nonminimal photon sector
of the full SME in Minkowski spacetime.
The relevant part of the Lagrange density
includes operators of arbitrary dimension $d$
that are both gauge invariant
and quadratic in the photon field $A_\mu$
\cite{2009akmmNonMin},
and these are listed in Table \PrNonMinPhotLagr.
This table is organized along the same lines
as Table \PrNonMinFermiLagr. 
The first column lists
the coefficient for Lorentz violation,
while the second column counts its independent components.
The next three columns provide
the corresponding operator appearing in the Lagrange density,
its mass dimension,
and the factor contracting the coefficient and the operator.
The last two columns list the properties of the operator
under CPT and Lorentz transformations,
using the same conventions as Table \PrNonMinFermiLagr. 

Three sections appear in Table \PrNonMinPhotLagr,
separated by horizontal lines.
The first section concerns the
conventional Lorentz-preserving Maxwell term
in the Lagrange density for the photon sector.
The second and third sections concern
coefficients associated with operators
of odd and even dimensions $d$, respectively.
Each of these sections has three rows
for the lowest three values of $d$,
along with a final row applicable
to the case of general $d$.
The notation for the coefficients incorporates
a superscript specifying the dimension $d$
of the corresponding operator.
Note that the mass dimension of the coefficients is $4-d$.
In each section,
the first row describes terms in the minimal SME,
and the match is provided
between the general notation for nonminimal coefficients
and the standard notation used for the minimal SME
in Table \PrQEDLagr.
In the case of mass dimension four,
there are 19 independent Lorentz-violating operators.
However,
for this case the number in the second column is listed as $19+1$
to allow for an additional Lorentz-preserving trace term,
which maintains consistency with the expression for general $d$
in the last row.

Table \PrNonMinPhotCoeffs\ summarizes
properties of spherical coefficients for Lorentz violation
in the nonminimal photon sector
of the full SME in Minkowski spacetime
\cite{2009akmmNonMin}.
The spherical coefficients
are combinations of the coefficients
listed in Table \PrNonMinPhotLagr\ that are of particular relevance
for observation and experiment.
They can be separated into two types.
One set consists of vacuum coefficients
that control leading-order effects on photon propagation
in the vacuum,
including birefringence and dispersion.
The complementary set contains the vacuum-orthogonal coefficients,
which leave photon propagation in the vacuum
unaffected at leading order.
The two parts of Table \PrNonMinPhotCoeffs\ reflect this separation,
with the part above the horizontal line
involving the vacuum coefficients
and the part below involving the vacuum-orthogonal ones.

In Table \PrNonMinPhotCoeffs,
the first column of the table
identifies the type of spherical coefficients,
while the second column lists the specific coefficient.
The coefficient notation
reflects properties of the corresponding operator.
Coefficients associated with operators leaving unchanged
the leading-order photon propagation in the vacuum
are distinguished by a negation diacritic $\neg$.
A symbol $k$ denotes coefficients for birefringent operators,
while $c$ denotes nonbirefringent ones.
The superscript $d$ refers to the operator mass dimension,
while the subscripts $n$, $j$, $m$
determine the frequency or wavelength dependence,
the total angular momentum,
and the $z$-component of the angular momentum,
respectively.
The superscripts $E$ and $B$ refer
to the parity of the operator,
while the numerals 0, 1, or 2
preceding $E$ or $B$ refer to the spin weight.
Note that the photon-sector coefficients in the minimal SME
correspond to the vacuum coefficients with $d=3,4$.
The third, fourth, and fifth columns
of Table \PrNonMinPhotCoeffs\ provide the allowed ranges of the dimension $d$
and of the indices $n$ and $j$.
The index $m$ can take values
ranging from $-j$ to $j$ in unit increments.
The final column gives the number
of independent coefficient components
for each operator of dimension $d$.

Table \PrNonMinFermPhot\ 
lists contributions to the Lagrange density 
of the nonminimal QED extension
for a single Dirac fermion in Minkowski spacetime
involving fermion and photon couplings with $d=5$ and 6
\cite{2016Ding,2018kl}.
The table is split into two parts,
one for the matter-photon interactions
and one for the four-fermion and the three-photon interactions.
The structure of both parts of this table 
follows that adopted for the quadratic terms in Table \PrNonMinFermiLagr,
with each row corresponding to a term in the Lagrange density. 
In a given row,
the first two entries list the coefficient for Lorentz violation
and the number of its independent components.
The third entry provides the operator,
the fourth indicates its mass dimension,
and the fifth is the numerical factor multiplying
the contraction of the coefficient and the operator
to give the term in the Lagrange density. 
The final two entries show the behavior
of the term under CPT and Lorentz transformations. 
Note that some of the terms with even $d$ 
include components that are Lorentz invariants. 
In this table,
parentheses enclosing a set of $n$ indices on covariant derivatives 
indicate symmetrization with a factor of $1/n!$.

The horizontal line in the first part of Table \PrNonMinFermPhot\ 
separates the entries according to their mass dimension $d=5$ and $d=6$.
For $d=5$,
two types of fermion-photon couplings exist.
One arises from 
using gauge-covariant derivatives $D_\mu$
instead of ordinary derivatives $\prt_\mu$
in the $d=5$ part of the Lagrange density for a free Dirac fermion
given in Table \PrNonMinFermiLagr.
The coefficients for Lorentz violation in the two theories 
can therefore be taken as identical.
The second type of coupling
is a gauge-nonminimal coupling 
between the field strength $F_{\mu\nu}$ and a fermion bilinear.
The nomenclature for the associated coefficient for Lorentz violation
is determined by the structure of the fermion bilinear
and includes a subscript $F$. 
For $d=6$,
the above two types of couplings also occur
but are accompanied by a third type
involving derivatives of $F_{\mu\nu}$ and a fermion bilinear.
The corresponding coefficients are distinguished
by the subscript $\prt F$.
The second part of Table \PrNonMinFermPhot\
includes the possible fermion self interactions,
which are quartic in the fermion fields,
and the sole photon self interaction for $d\leq 6$,
which is cubic in the photon field.
None of the operators listed in Table \PrNonMinFermPhot\ 
are of renormalizable dimension.

Table \PrNonMinNuCoeffs\ provides a guide to spherical coefficients for Lorentz violation
in the neutrino sector of the full SME
\cite{2011akmm}.
Use of these coefficients simplifies
analyses in many circumstances
because their rotation properties are comparatively simple.
The table is separated by horizontal lines
into seven sections,
reflecting the various theoretical scenarios
listed in the first column.
The spherical coefficients relevant to each scenario
are given in the second column.
The third column shows the allowed values
of the corresponding operator dimensions,
while the fourth column provides
the permissible range of the angular-momentum quantum number $j$.
The last two columns
display the number of independent coefficients involved
and the CPT properties of the corresponding operators.

The effective hamiltonian governing
the propagation and mixing of neutrinos
in the presence of Lorentz-violating operators
of arbitrary dimensions
contains four sets of effective spherical coefficients,
listed in the first section
of Table \PrNonMinNuCoeffs.
The observables in a given experiment lie among
these effective coefficients.
In the leading-order relativistic limit,
they are formed from combinations of
the ten sets of basic spherical coefficients
displayed in the second section of the table.
Six of these basic sets involve Dirac-type operators
coupling neutrinos to neutrinos
and antineutrinos to antineutrinos,
while four involve Majorana-type operators
coupling neutrinos to antineutrinos.
The remainder of the sections in the table
concern various limiting cases of the general formalism.
The spherical coefficients
associated with operators of renormalizable dimension
are shown in the third section.
Their relation to the cartesian coefficients
listed in Table \PrNuCoeffs\ is given in Table \PrNuDefs.
The fourth section of Table \PrNonMinNuCoeffs\ contains the coefficients appearing
in the massless limit of the SME.
The remaining three sections concern
the special limits
of flavor-blind and oscillation-free operators,
of simultaneously diagonalizable operators,
and of various types of isotropic operators.
Constraints on many of the coefficients
listed in these sections
have been obtained and are given in the data tables.

\subsection{D.\ \ Nonminimal sectors, Riemann spacetime}

Tables \PrSMEGravityGaugeLag--\PrSMEYukawaGravityGaugeLag\
provide the explicit forms of all terms 
with $d\leq 6$ and without background derivatives
in the Lagrange density $\cL_{\rm SME}$ in Riemann spacetime
\cite{kl21-1}.
It is convenient to express this Lagrange density as a sum
of contributions from different sectors,
\bea
\cL_{\rm SME} &=&
\cL_{\rm gravity} 
+ \cL_{\rm gauge} 
+\cL_{\rm lepton} 
+ \cL_{\rm quark} 
\nonumber\\ &&
+ \cL_{\rm Higgs} 
+ \cL_{\rm Yukawa} ,
\nonumber
\eea
where $\cL_{\rm gravity}$ involves only gravitational fields,
$\cL_{\rm gauge}$ also incorporates gauge fields,
$\cL_{\rm lepton}$ adds leptons, and so on.
The tables display each component Lagrange density
prior to the breaking of the gauge group SU(3)$\times$SU(2)$\times$U(1),
using the same notation and conventions as 
Tables \PrSMEFermionLagr\ and \PrSMEBosonLagr.
In each of the Tables \PrSMEGravityGaugeLag--\PrSMEYukawaGravityGaugeLag,
the first line represents the usual theory coupling 
General Relativity to the Standard Model,
while the other lines contain terms 
violating local Lorentz invariance, diffeomorphism invariance, or both.
Other terms appearing in
Tables \PrSMEGravityGaugeLag--\PrSMEYukawaGravityGaugeLag\
are formed as contractions of a dynamical operator
$\cO_{\mu\ldots}{}^{\nu\ldots}{}_{a\ldots}$
with a breve coefficient
$\uk^{\mu\cdots}{}_{\nu\cdots}{}^{a\cdots}$.
For the latter,
the breve diacritic is a convenient notation denoting a linear combination of 
arbitrary couplings of the vierbien, metric, and Levi-Civita tensor
to background fields.
For instance,
a breve coefficient $\uk^\mu$ with a single contravariant spacetime index
can be expanded as 
$\uk^\mu = k^\mu + k_\nu g^{\mu\nu} + k^a e^\mu{}_a + \ldots$,
thereby generating a series of terms in the Lagrange density
involving the coefficients $k^\mu$, $k_\nu$, $k^a$, $\ldots$.

Table \PrSMEGravityGaugeLag\ lists the terms in
$\cL_{\rm gravity}$ and $\cL_{\rm gauge}$.
The first column lists the component 
$\cL_{\rm gravity}^{(d)}$ or $\cL_{\rm gauge}^{(d)}$
containing operators of specified mass dimension $d$,
while the second column contains the explicit form 
of the corresponding terms. 
Each breve coefficient $\uk$ is real,
and its index symmetry inherited 
from that of its associated dynamical operator.
In the gauge sector,
a trace $\tr(\cO)$ of an operator $\cO$ 
is understood to be taken in the representation of $\cO$.
Table \PrSMEGravityGaugeLag\ incorporates also 
terms with dynamical operators that are total derivatives,
which are invariant under general coordinate transformations
only up to a surface term.
In particular,
the operators contracted with the breve coefficients
$(\uk^{(5)}_{{\rm CS},1})_\mu$
and $(\uk^{(5)}_{{\rm CS},2})_\mu$
are Chern-Simons terms expressed in terms of the vierbein and spin connection.

Terms in the Lagrange density involving leptons and quarks,
along with their couplings to the gravitational and gauge fields,
are displayed in Table \PrSMELeptonQuarkGravityGaugeLag.
The first column presents the various components of 
$\cL_{\rm lepton}+\cL_{\rm quark}$ at fixed $d$,
first listing components without quark fields,
then components without lepton fields,
and finally the component $\cL^{(6)}_{{\rm quark,lepton}}$
that includes four-point interactions between leptons and quarks.
The notation for the breve coefficients follows standard conventions
in the literature,
with backgrounds of differing spin and CPT properties
represented by different Latin letters.
The breve coefficients carry flavor indices,
so flavor violations are also incorporated. 
The index symmetry of each breve coefficient
is fixed by that of its dynamical operator.
Where the symbol $\hc$ for the hermitian conjugate appears
in a specific component of the Lagrange density,
it applies to all the terms in that component.
If the symbol $\hc$ is absent,
all breve coefficients in that component
can be taken as hermitian in generation space.

Table \PrSMEHiggsGravityGaugeLag\
provides all terms for the Lagrange density
of the components $\cL_{\rm Higgs}$ in the Higgs sector,
including couplings to gravitational and gauge fields.
The first column lists 
the components $\cL^{(d)}_{\rm Higgs}$ for each $d$,
while the second column provides the explicit form of the corresponding terms. 
All the breve coefficients $\uk$ can be taken as real,
and the index symmetry of each matches that 
of the corresponding dynamical operator.
Terms involving breve coefficients without spacetime indices
represent either scalar coupling constants
or effects dependent on spacetime position.
Note that the scalar coupling constants of this type in 
$\cL^{(2)}_{\rm Higgs}$, $\cL^{(4)}_{\rm Higgs}$
and the scalar coupling constant 
in the trace piece of the breve coefficient $(\uk_{\ph\ph})^{\mu\nu}$
can be viewed as renormalizations 
of the scalar couplings in the usual Lagrange density $\cL_{\rm Higgs,0}$ 
for the Higgs field.

Table \PrSMEYukawaGravityGaugeLag\
contains all terms generalizing the usual Yukawa couplings
together with their couplings to gravitational and gauge fields.
The first column lists the various component Lagrange densities
containing operators at the specified $d$,
while the second column shows the explicit forms
of the corresponding terms.
The notation for the breve coefficients
matches established conventions in the literature.
As before,
the index symmetry of each breve coefficient
is fixed by that of the associated dynamical operator.
The symbol $\hc$ for the hermitian conjugate is understood
to apply to all the terms for that component Lagrange density.
For components lacking the symbol $\hc$,
all the breve coefficients can be taken as hermitian in generation space.
Note that the three first entries for $\cL^{(4)}_{\rm Yukawa}$ 
include scalar coupling constants that can be interpreted
as renormalizing the usual Yukawa couplings
in the Lagrange density $\cL_{\rm Yukawa,0}$.

\subsection{E.\ \ Linearized gravity}

Table \PrLinGrav\ lists the operators contributing
to the quadratic Lagrange density of the pure-gravity sector
in the linearized limit 
\cite{2017km}.
The linearization assumes the metric $g_{\mn}$
is expanded about a Minkowski background,
$g_\mn = \et_\mn+ h_\mn$.
Only quadratic terms in $h_\mn$ are kept,
so each term of dimension $d\geq 2$ in the Lagrange density has the form 
$h_{\mu\nu} \widehat\cK^{\mu\nu\rh\si} h_{\rh\si}/4$,
where $\widehat\cK^{\mu\nu\rh\si}$ 
is a generic operator formed as the product of a coefficient 
$\cK^{\mu\nu\rh\si\ve_1\ve_2\ldots\ve_{d-2}}$
of mass dimension $4-d$ with a factor 
$\prt_{\ve_1} \prt_{\ve_2}\ldots \prt_{\ve_{d-2}}$
containing $d-2$ derivatives.
For these operators,
it is convenient to denote an index contracted into a derivative 
by a circle index $\circ$,
with $n$-fold contractions denoted by $\circ^n$,
so that
$\widehat\cK^{\mu\nu\rh\si} = \cK^{\mu\nu\rh\si\circ^{d-2}}$.

The operators for linearized gravity
can be classified according to the index symmetries
of the corresponding coefficients,
which distinguishes 14 possibilities
denoted as shown in the first column of Table \PrLinGrav.
The 14 operators fall naturally into three classes,
separated by horizontal lines in the table 
and determined by the symmetry of the first four indices 
on each operator.
The second entry in each row displays
the complete index symmetry of the operator
using Young tableaux
\cite{1962mh}.
The third entry provides the number of independent components
of the corresponding coefficient,
while the fourth entry lists the allowed values of $d$.
The fifth entry shows the CPT handedness of the operator.
The operators listed yield 
all possible quadratic terms in the Lagrange density,
irrespective of their transformation properties 
under Lorentz transformations or under linearized diffeomorphisms.
The latter are the standard gauge transformations of linearized gravity,
$h_{\mu\nu} \to h_{\mu\nu} + \prt_\mu \xi_\nu + \prt_\nu \xi_\mu$,
which depend on four parameters $\xi_\mu$.
The Lorentz-invariant terms are constructed from operators
involving complete traces over the coefficient indices,
while the three gauge-invariant operators are identified
in the last column of Table \PrLinGrav.
Note that the usual Lorentz- and gauge-invariant 
linearized Einstein-Hilbert Lagrange density
is contained as a special case of the operator in the first row.
Also, 
certain limits of the 11 gauge-violating operators
may preserve a subset of the four gauge symmetries,
but they always break at least one.

Table \PrLinGravSpher\ lists spherical coefficients
for some key gauge-invariant terms
in the Lagrange density for the pure-gravity sector
in the linearized limit
\cite{2016Kostelecky}.
These spherical coefficients are
special combinations of coefficients
appearing in Table \PrLinGrav\ that turn out to be 
of direct relevance to studies of gravitational waves
and laboratory experiments on Newton gravity.
The vacuum coefficients shown in Table \PrLinGravSpher\ 
control leading-order modifications 
to the propagation of gravitational waves,
including birefringent, dispersive, and anisotropic effects.
The Newton coefficients govern corrections to Newton gravity,
including changes to the inverse-square law and noncentral forces.  

In Table \PrLinGravSpher,
the first column states the type of spherical coefficients,
and the second column identifies each coefficient
using notation that specifies the mass dimension $d$
of the corresponding operator via a superscript
and the angular-momentum properties by subscripts $j$ and $m$.
The allowed range of the dimension $d$
and the angular-momentum index $j$ 
are displayed in the third and fourth columns,
respectively.
The index $m$ takes integer values ranging from $-j$ to $j$,
as usual.
Note that the notation for the vacuum spherical coefficients 
for linearized gravity
parallels that for the vacuum spherical coefficients
in the photon sector,
but the allowed ranges differ
due to the different spins of the fields. 
The final column of the table
lists the number of independent components 
of each spherical coefficient at fixed $d$.

Table \PrLinGravDef\ provides definitions
for some combinations of coefficients
forming part of the gauge-invariant sector
of the Lagrange density for pure gravity in the linearized limit
\cite{2016Kostelecky}.
The table is split by horizontal lines into five sections.
The first gives the definition 
of the two-index dual cartesian coefficient $\std{4}_{\ka\la}$
in terms of the six-index cartesian coefficient  
$ s^{(4)}{}^{\mu\rh\al\nu\si\be}$ 
appearing in Table \PrLinGrav. 
This dual cartesian coefficient plays a central role
in linearized gravity
because it contains the usual linearized Einstein-Hilbert term
and because its effects mimic those of a modified metric.
The second section of the table
expresses the nine vacuum spherical coefficients $k^{(4)}_{(I)jm}$
in terms of components of the dual cartesian coefficients.
The third section provides the definition
linking two alternative forms of spherical coefficients
found in the literature,
which differ by a factor and a phase.
The remaining two sections present relationships
between the Newton spherical coefficients with $d=4$ and $d=6$
and the cartesian and vacuum spherical coefficients.
These expressions permit the comparison 
of constraints obtained from studies of gravitational waves
and from Newton gravity. 

Tables \PrFermionGravityLag--\PrFermionGravityNRcoeffs\
display information relevant 
to the fermion-gravity couplings involving linearized gravity.
Contributions to the Lagrange density containing 
operators of mass dimension $d\leq 5$
are presented in Table \PrFermionGravityLag,
allowing for operators constructed from a massive Dirac fermion $\ps$
and up to one power of the metric fluctuation $h_{\mu\nu}$.
The first column of this table lists 
possible pieces of the Lagrange density,
with each defined according to its operator mass dimension 
and the number of derivatives of $h_{\mu\nu}$ that appear. 
Any given piece is a sum of component terms,
forming the combinations listed in the second column of the table.
The first two rows of the table display
standard expressions in linearized fermion-gravity theory,
which are Lorentz and gravitational-gauge invariant.
Other terms in the table typically are Lorentz violating 
and gravitational-gauge violating,
with exceptions occurring only for special coefficients.
The labeling of the coefficients appearing in all these terms 
matches standard usage in the literature.
It distinguishes spin and CPT properties,
and the superscript L indicates the linearized nature 
of the corresponding operators.

Certain combinations of the linearized coefficients 
appearing in Table \PrFermionGravityLag\
frequently appear in experimental analyses.
These combinations are denoted by tilde coefficients,
and they are defined in Table \PrFermionGravityTilde.
They can be viewed as generalizations to linearized gravity
of the definitions for Minkowski spacetime
introduced in Table \PrFermionTilde. 
The first column of Table \PrFermionGravityLag\ lists the tilde coefficients,
while the second column presents the corresponding combinations
of linearized coefficients. 
The last column gives the number of independent components
contained in each tilde coefficient.
The table assumes a generic fermion of mass $m$,
so 44 combinations of linearized coefficients
can in principle be measured for each flavor of fermion.
The tilde coefficients are expressed in the Sun-centered frame,
with $J$, $K$, $L$ ranging over $X$, $Y$, $Z$
and the repeated index $\Si\Si$
denoting summation over indices $TT$, $XX$, $YY$, $ZZ$.

In the nonrelativistic and weak-gravity limit,
the hamiltonian governing fermion-gravity physics
contains special combinations of linearized coefficients,
called nonrelativistic coefficients and denoted with superscripts NR. 
Table \PrFermionGravityNRcoeffs\ presents the nonrelativistic coefficients
relevant for laboratory experiments involving a uniform gravitational field.
The first column of the table shows the nonrelativistic coefficients,
while the second column displays their definitions
in terms of the linearized coefficients
appearing in Table \PrFermionGravityLag.
The indices $j$, $k$, $l$, $m$, $n$ 
range over the three spatial cartesian coordinates
chosen for the laboratory analysis,
while the temporal index is denoted as $t$.
The repeated index $ss$
indicates summation over the temporal index $tt$
and all the spatial indices $jj$.

\section*{Acknowledgments}

This work was supported in part
by the U.S.\ Department of Energy under grant {DE}-SC0010120
and by the Indiana University Center for Spacetime Symmetries.

\onecolumngrid

\newpage

	\begin{center}									
	\scalebox{0.95}{									
 \\[2pt]		
	$		|a^{\rm NR}_{200}|,\ |c^{\rm NR}_{200}|		$&$	<2.1 \times10^{-2} \gev^{-1}	$&		H, ${\ol {\rm H}}$ spectroscopy		&	\rr{2024vargas}*	\\	[2pt]	
	$		|a^{\rm NR}_{220}|,\ |c^{\rm NR}_{220}|		$&$	<3.0 \times10^{-1} \gev^{-1}	$&	\dit{	H, ${\ol {\rm H}}$ spectroscopy	}	&	\rr{2024vargas}*	\\	[2pt]	
	$		|\HzBnr{210}|,\ |\gzBnr{210}|		$&$	<1.0 \times10^{-2} \gev^{-1}	$&	\dit{	H, ${\ol {\rm H}}$ spectroscopy	}	&	\rr{2024vargas}*	\\	[2pt]	
	$		|\HoBnr{210}|,\ |\goBnr{210}|		$&$	<5.2 \times10^{-2} \gev^{-1}	$&	\dit{	H, ${\ol {\rm H}}$ spectroscopy	}	&	\rr{2024vargas}*	\\	[2pt]	
	$		|\HzBnr{230}|,\ |\gzBnr{230}|		$&$	<4.6\times10^{-1} \gev^{-1}	$&	\dit{	H, ${\ol {\rm H}}$ spectroscopy	}	&	\rr{2024vargas}*	\\	[2pt]	
	$		|\HoBnr{230}|,\ |\goBnr{230}|		$&$	<2.8\times10^{-1} \gev^{-1}	$&	\dit{	H, ${\ol {\rm H}}$ spectroscopy	}	&	\rr{2024vargas}*	\\	[2pt]	
	$		|\HzBnr{210}|,\ |\gzBnr{210}|		$&$	<8.4 \times10^{-11} \gev^{-1}	$&		H beam		&	\rr{2024nowak}	\\	[2pt]	
	$		|\HoBnr{210}|,\ |\goBnr{210}|		$&$	<4.2 \times10^{-11} \gev^{-1}	$&	\dit{	H beam	}	&	\rr{2024nowak}	\\	[2pt]	
	$		|\Re\HzBnr{211}|,\ |\Im\HzBnr{211}|,\ |\Re\gzBnr{211}|,\ |\Im\gzBnr{211}|		$&$	<7 \times10^{-16} \gev^{-1}	$&		H maser		&	\rr{2015akav}*	\\[2pt]		
	$		|\Re\HoBnr{211}|,\ |\Im\HoBnr{211}|,\ |\Re\goBnr{211}|,\ |\Im\goBnr{211}|		$&$	<4 \times10^{-16} \gev^{-1}	$&	\dit{	H maser	}	&	\rr{2015akav}*	\\[2pt]		
			\hlinetwo												
	$		|a^{(5)TTX}_{\rm eff}|		$&$	< 3.4\times 10^{-8}\gev^{-1}	$&		1S-2S transition		&	\rr{2015akav}*	\\[2pt]		
	$		|a^{(5)TTY}_{\rm eff}|		$&$	< 5.6\times 10^{-8}\gev^{-1}	$&	\dit{	1S-2S transition	}	&	\rr{2015akav}*	\\[2pt]		
	$		|a^{(5)TTZ}_{\rm eff}|		$&$	< 1.3\times 10^{-7}\gev^{-1}	$&	\dit{	1S-2S transition	}	&	\rr{2015akav}*	\\[2pt]		
	$		|a^{(5)KKX}_{\rm eff}|		$&$	< 6.7\times 10^{-8}\gev^{-1}	$&	\dit{	1S-2S transition	}	&	\rr{2015akav}*	\\[2pt]		
	$		|a^{(5)KKY}_{\rm eff}|		$&$	< 1.1\times 10^{-7}\gev^{-1}	$&	\dit{	1S-2S transition	}	&	\rr{2015akav}*	\\[2pt]		
	$		|a^{(5)KKZ}_{\rm eff}|		$&$	< 2.5\times 10^{-7}\gev^{-1}	$&	\dit{	1S-2S transition	}	&	\rr{2015akav}*	\\[2pt]		
	$		\ring{m}_0,\ \ring{m}_2		$&$	< 1 \times 10^{-18} \gev^{-1}	$&		Astrophysics		&	\rr{2025Petrov}*	\\[2pt]		
	$		\ring{a}_0,\ \ring{a }_2		$&$	< 3 \times 10^{-28} \gev^{-1}	$&	\dit{	Astrophysics	}	&	\rr{2025Petrov}*	\\[2pt]		
			\hlinethree												
			\end{tabular}												
			} \end{center}												
															
			\begin{center}												
			\scalebox{0.95}{												
	\\	[14pt]	
	$		(a_L)_{e\ta}^T		$&$		<6.0 \times 10^{-13} \gev		$&		JUNO		&		\rr{2025Araya}*		\\		
	$		|\aeff{3}{00}{e\ta}|		$&$		<6.5\times 10^{-20}\gev		$&		SNO		&		\rr{2018SNO}		\\		
	$		|\aeff{3}{10}{e\ta}|		$&$		<2.8\times 10^{-21}\gev		$&	\dit{	SNO	}	&		\rr{2018SNO}		\\		
	$		|\re\aeff{3}{11}{e\ta}|		$&$		<1.5\times 10^{-21}\gev		$&	\dit{	SNO	}	&		\rr{2018SNO}		\\		
	$		|\im\aeff{3}{11}{e\ta}|		$&$		<1.7\times 10^{-21}\gev		$&	\dit{	SNO	}	&		\rr{2018SNO}		\\	[6pt]	
	$		(a_R)_{\bar e \bar \ta}^X		$&$		(-5.6 \pm 8.0) \times 10^{-20} \gev		$&		Daya Bay		&		\rr{2018DayaBay}		\\		
	$		(a_R)_{\bar e \bar \ta}^Y		$&$		(-0.9 \pm 8.0) \times 10^{-20} \gev		$&	\dit{	Daya Bay	}	&		\rr{2018DayaBay}		\\		
	$		\re(a^T_{e\ta})		$&$		< 4.1\times 10^{-23}\gev		$&		Super-Kamiokande		&		\rr{2014SuperK}		\\		
	$		\im(a^T_{e\ta})		$&$		< 2.8\times 10^{-23}\gev		$&	\dit{	Super-Kamiokande	}	&		\rr{2014SuperK}		\\		
	$		|\Re(a_L)^T_{e\ta}|, \ |\Im(a_L)^T_{e\ta}|		$&$		<7.8\times 10^{-20}\gev		$&		Double Chooz		&		\rr{2013KatoriChooz}		\\		
	$		|\Re(a_L)^X_{e\ta}|, \ |\Im(a_L)^X_{e\ta}|		$&$		<4.4\times 10^{-20}\gev		$&	\dit{	Double Chooz	}	&		\rr{2013KatoriChooz}		\\		
	$		|\Re(a_L)^Y_{e\ta}|, \ |\Im(a_L)^Y_{e\ta}|		$&$		<9.0\times 10^{-20}\gev		$&	\dit{	Double Chooz	}	&		\rr{2013KatoriChooz}		\\		
	$		|\Re(a_L)^Z_{e\ta}|, \ |\Im(a_L)^Z_{e\ta}|		$&$		<2.7\times 10^{-19}\gev		$&	\dit{	Double Chooz	}	&		\rr{2013KatoriChooz}		\\		
	$	\brry{l}	\big|-(a_L)_{e\ta}^T - 0.29(a_L)_{e\ta}^Z+ 0.0042[-1.46(c_L)_{e\ta}^{TT} \\ \hspace{8em} -0.57(c_L)_{e\ta}^{TZ}+0.38(c_L)_{e\ta}^{ZZ}]\gev\big|	\hspace{2em}\erry	$&$	\trry{	<7.8\times 10^{-20}\gev	}	$&	\begin{tabular}{c}\dit{	Double Chooz	}\\ \pt Q \end{tabular}	&	\begin{tabular}{c}	\rr{2012DChooz}	\\ \pt Q \end{tabular}	\\		
	$	\brry{l}	\big|-0.91(a_L)_{e\ta}^X+0.29(a_L)_{e\ta}^Y+0.0042[-1.83(c_L)_{e\ta}^{TX} \\ \qquad +0.58(c_L)_{e\ta}^{TY}-0.52(c_L)_{e\ta}^{XZ}+0.16(c_L)_{e\ta}^{YZ}]\gev\big|	\hspace{1.8em}\erry	$&$	\trry{	<6.6\times 10^{-20}\gev	}	$&	\begin{tabular}{c}\dit{	Double Chooz	}\\ \pt Q \end{tabular}	&	\begin{tabular}{c}	\rr{2012DChooz}	\\ \pt Q \end{tabular}	\\		
	$	\brry{l}	\big|0.29(a_L)_{e\ta}^X+0.91(a_L)_{e\ta}^Y+0.0042[0.58(c_L)_{e\ta}^{TX} \\ \qquad +1.83(c_L)_{e\ta}^{TY}+0.16(c_L)_{e\ta}^{XZ}+0.52(c_L)_{e\ta}^{YZ}]\gev\big|	\hspace{1.8em}\erry	$&$	\trry{	<7.0\times 10^{-20}\gev	}	$&	\begin{tabular}{c}\dit{	Double Chooz	}\\ \pt Q \end{tabular}	&	\begin{tabular}{c}	\rr{2012DChooz}	\\ \pt Q \end{tabular}	\\		
	$		\big|0.0042[0.26((c_L)_{e\ta}^{XX}-(c_L)_{e\ta}^{YY})+0.75(c_L)_{e\ta}^{XY}]\gev\big|	\hspace{2em}	$&$		<5.4\times 10^{-20}\gev		$&	\dit{	Double Chooz	}	&		\rr{2012DChooz}		\\		
	$		\big|0.0042[0.38((c_L)_{e\ta}^{XX}-(c_L)_{e\ta}^{YY})-0.53(c_L)_{e\ta}^{XY}]\gev\big|	\hspace{2em}	$&$		<5.4\times 10^{-20}\gev		$&	\dit{	Double Chooz	}	&		\rr{2012DChooz}		\\	[14pt]	
	$		|(a_L)_{\mu\mu}^T|		$&$		<2 \times 10^{-7} \gev		$&		Leptons, SU(2)$_{\rm L}$ invariance		&		\rr{2020Crivellin}*		\\		
	$		|(a_L)_{\mu\mu}^X|		$&$		<4 \times 10^{-23} \gev		$&	\dit{	Leptons, SU(2)$_{\rm L}$ invariance	}	&		\rr{2020Crivellin}*		\\		
	$		|\aeff{3}{00}{\mu\mu}|		$&$		<8.2\times 10^{-20}\gev		$&		SNO		&		\rr{2018SNO}		\\		
	$		|\aeff{3}{10}{\mu\mu}|		$&$		<5.4\times 10^{-21}\gev		$&	\dit{	SNO	}	&		\rr{2018SNO}		\\		
	$		|\re\aeff{3}{11}{\mu\mu}|		$&$		<2.9\times 10^{-21}\gev		$&	\dit{	SNO	}	&		\rr{2018SNO}		\\		
	$		|\im\aeff{3}{11}{\mu\mu}|		$&$		<3.2\times 10^{-21}\gev		$&	\dit{	SNO	}	&		\rr{2018SNO}		\\	[6pt]	
	$		(a_R)_{\bar \mu \bar \mu}^X		$&$		(9 \pm 45) \times 10^{-20} \gev		$&		Daya Bay		&		\rr{2018DayaBay}		\\		
	$		(a_R)_{\bar \mu \bar \mu}^Y		$&$		(-9 \pm 45) \times 10^{-20} \gev		$&	\dit{	Daya Bay	}	&		\rr{2018DayaBay}		\\		
	$		|(a_L)^0_{\mu\mu}-(a_L)^0_{\ta\ta}|		$&$		<1.9\times 10^{-23}\gev		$&		Super-Kamiokande		&		\rr{2018Barenboim}*		\\		
	$		|(a_L)^X_{\mu\mu}|, \ |(a_L)^Y_{\mu\mu}||		$&$		< 4.8 \times 10^{-20} \gev		$&		T2K		&		\rr{2016Quilain}		\\		
	$		|(a_L)^0_{\mu\mu}-(a_L)^0_{\ta\ta}|		$&$		<6.3\times 10^{-24}\gev		$&		Super-Kamiokande		&		\rr{2015Diaz}*		\\		
			\hlinethree																
			\end{tabular}}																
			\end{center}

			\begin{center}\scalebox{0.95}{																
 \\	
	$		\sum_{d} (d-3) (17\gev)^{d-4} (\acfc{d}{} - \ccfc{d}{})		$&$	2.37 \pm 0.32^{+0.34}_{-0.24} \times 10^{-5}	$&		OPERA time of flight		&	\rr{2011akmm}*	\\	
	$		\sum_{d} (d-3) (3\gev)^{d-4} (\acfc{d}{} - \ccfc{d}{})		$&$	5.1\pm 2.9 \times 10^{-5}	$&		MINOS time of flight		&	\rr{2011akmm}*	\\	
	$		\big|\sum_{d} (d-3) (30\gev)^{d-4} (\pm\acfc{d}{} - \ccfc{d}{})\big|		$&$	< 4 \times 10^{-5}	$&		Fermilab time of flight		&	\rr{2011akmm}*	\\	
	$		\big|\sum_{d} (d-3) (30\gev)^{d-4}\, \acfc{d}{} \big|		$&$	< 3.5 \times 10^{-5}	$&		Fermilab $\nu\ol\nu$ comparison		&	\rr{2011akmm}*	\\	
	$		\big|\sum_{d} (d-3) (10~{\rm MeV})^{d-4}\, (\acfc{d}{} + \ccfc{d}{})\big|		$&$	< 2 \times 10^{-9}	$&		SN1987A time of flight		&	\rr{2011akmm}*	\\	
			\hlinethree											
			\end{tabular}}											
			\end{center}											

			\begin{center}\scalebox{0.95}{												
\begin{tabular}{@{}cr@{\hspace{12pt}}ll@{}}												
			\multicolumn{4}{c}{Table \DaQuarkDimThree.\ Quark sector, $d = 3$}\\												
			\hlineone												
			{\bf Combination}		&	\multicolumn{1}{c}{{\bf Result}}	&		\multicolumn{1}{c}{{\bf System}}		&	\multicolumn{1}{c}{{\bf Ref.}}	\\		
			\hlinetwo												
	$		\De a^K_0		$&$	(-6.0\pm 7.7_{\rm stat}\pm 3.1_{\rm syst}) \times 10^{-18} \gev	$&		$K$ oscillations		&	\rr{2013DeSantis}	\\	[2pt]	
	$		\De a^K_X		$&$	(0.9\pm 1.5_{\rm stat}\pm 0.6_{\rm syst}) \times 10^{-18} \gev	$&	\dit{	$K$ oscillations	}	&	\rr{2013DeSantis}	\\	[2pt]	
	$		\De a^K_Y		$&$	(-2.0\pm 1.5_{\rm stat}\pm 0.5_{\rm syst}) \times 10^{-18} \gev	$&	\dit{	$K$ oscillations	}	&	\rr{2013DeSantis}	\\	[2pt]	
	$		\De a^K_Z		$&$	(3.1\pm 1.7_{\rm stat}\pm 0.5_{\rm syst}) \times 10^{-18} \gev	$&	\dit{	$K$ oscillations	}	&	\rr{2013DeSantis}	\\	[2pt]	
	$		\De a^K_0		$&$	(0.4 \pm 1.8) \times 10^{-17} \gev	$&	\dit{	$K$ oscillations	}	&	\rr{2008KLOE}, \rr{DiDomenico}	\\	[2pt]	
	$		\De a^K_X		$&$	(-6.3 \pm 6.0) \times 10^{-18} \gev	$&	\dit{	$K$ oscillations	}	&	\rr{2008KLOE}	\\
\end{tabular}}												
			\end{center}

			\begin{center}\scalebox{0.95}{												
			\begin{tabular}{@{}cr@{\hspace{12pt}}ll@{}}												
			\multicolumn{4}{c}{Table \DaQuarkDimFour.\ Quark sector, $d=4$ (part 1 of 4)}\\												
			\hlineone												
			{\bf Combination}		&	\multicolumn{1}{c}{{\bf Result}}	&		\multicolumn{1}{c}{{\bf System}}		&	\multicolumn{1}{c}{{\bf Ref.}}	\\		
			\hlinetwo												
	$		c_u^{TX}		$&$	(-25\mbox{ to }6.6)\times 10^{-5}	$&		Deep inelastic scattering		&	\rr{2022ZEUS}	\\	[2pt]	
	$		c_u^{TY}		$&$	(-17\mbox{ to }9.8)\times 10^{-5}	$&	\dit{	Deep inelastic scattering	}	&	\rr{2022ZEUS}	\\	[2pt]	
	$		c_u^{XY}		$&$	(-32\mbox{ to }4.1)\times 10^{-5}	$&	\dit{	Deep inelastic scattering	}	&	\rr{2022ZEUS}	\\	[2pt]	
	$		c_u^{XZ}		$&$	(-5.4\mbox{ to }1.4)\times 10^{-4}	$&	\dit{	Deep inelastic scattering	}	&	\rr{2022ZEUS}	\\	[2pt]	
	$		c_u^{YZ}		$&$	(-3.7\mbox{ to }2.1)\times 10^{-4}	$&	\dit{	Deep inelastic scattering	}	&	\rr{2022ZEUS}	\\	[2pt]	
	$		c_u^{XX}- c_d^{YY}		$&$	(-2.1\mbox{ to }2.5)\times 10^{-4}	$&	\dit{	Deep inelastic scattering	}	&	\rr{2022ZEUS}	\\
\end{tabular}}												
			\end{center}

			\begin{center}\scalebox{0.95}{												
			\begin{tabular}{@{}cr@{\hspace{12pt}}ll@{}}												
			\multicolumn{4}{c}{Table \DaQuarkDimFour.\ Quark sector, $d=4$ (part 2 of 4)}\\												
			\hlineone												
			{\bf Combination}		&	\multicolumn{1}{c}{{\bf Result}}	&		\multicolumn{1}{c}{{\bf System}}		&	\multicolumn{1}{c}{{\bf Ref.}}	\\		
			\hlinetwo												
	$		c_{u}^{TT} 		$&$	(-4.3\mbox{ to }1.4)\times 10^{-2}	$&		Drell-Yan process		&	\rr{2024lunghi}*	\\	[2pt]	
	$		c_{u}^{ZZ} 		$&$	(-4.3\mbox{ to }13)\times 10^{-4}	$&	\dit{	Drell-Yan process	}	&	\rr{2024lunghi}*	\\	[2pt]	
	$		d_{u}^{TT} 		$&$	(-5.6\mbox{ to }1.8)\times 10^{-2}	$&	\dit{	Drell-Yan process	}	&	\rr{2024lunghi}*	\\	[2pt]	
	$		d_{u}^{ZZ} 		$&$	(-5.6\mbox{ to }17)\times 10^{-4}	$&	\dit{	Drell-Yan process	}	&	\rr{2024lunghi}*	\\	[2pt]	
	$		c_{d}^{TT} 		$&$	<6.6\times 10^{-1} 	$&	\dit{	Drell-Yan process	}	&	\rr{2024lunghi}*	\\	[2pt]	
	$		c_{d}^{ZZ} 		$&$	(-1.9\mbox{ to }5.9)\times 10^{-2}	$&	\dit{	Drell-Yan process	}	&	\rr{2024lunghi}*	\\
\end{tabular}}												
			\end{center}												
															
			\begin{center}\scalebox{0.95}{												
			\begin{tabular}{@{}cr@{\hspace{12pt}}ll@{}}												
			\multicolumn{4}{c}{Table \DaQuarkDimFour.\ Quark sector, $d=4$ (part 3 of 4)}\\												
			\hlineone												
			{\bf Combination}		&	\multicolumn{1}{c}{{\bf Result}}	&		\multicolumn{1}{c}{{\bf System}}		&	\multicolumn{1}{c}{{\bf Ref.}}	\\		
			\hlinetwo												
	$		\ring{c}^u - \frac 34 \tilde\kappa_{\rm tr} - \frac 38 (\ring{f}^u)^2		$&$	> -3\times10^{-23}	$&		Astrophysics		&	\rr{2017Schreck}*	\\	[2pt]	
	$		|\ring{d}^u|		$&$	< 3\times10^{-23}	$&	\dit{	Astrophysics	}	&	\rr{2017Schreck}*	\\	[2pt]	
	$		|\ring{e}^u|		$&$	< 9\times10^{-12}	$&	\dit{	Astrophysics	}	&	\rr{2017Schreck}*	\\	[2pt]	
	$		|\ring{g}^u|		$&$	< 9\times10^{-12}	$&	\dit{	Astrophysics	}	&	\rr{2017Schreck}*	\\	[2pt]	
	$		c_{33}^{(4)u}- \frac 38 (f_3^{(4)u})^2		$&$	> -6\times10^{-22}	$&	\dit{	Astrophysics	}	&	\rr{2017Schreck}*	\\	[2pt]	
	$		|d_{33}^{(4)u}|		$&$	< 6\times10^{-22}	$&	\dit{	Astrophysics	}	&	\rr{2017Schreck}*	\\
\end{tabular}}												
			\end{center}

			\begin{center}\scalebox{0.95}{												
			\begin{tabular}{@{}cr@{\hspace{12pt}}ll@{}}												
			\multicolumn{4}{c}{Table \DaQuarkDimFour.\ Quark sector, $d=4$ (part 4 of 4)}\\												
			\hlineone												
			{\bf Combination}		&	\multicolumn{1}{c}{{\bf Result}}	&		\multicolumn{1}{c}{{\bf System}}		&	\multicolumn{1}{c}{{\bf Ref.}}	\\		
			\hlinetwo												
	$		c_t^{\mu\nu}		$&$	< 10^{-4}	$&		Astrophysics		&	\rr{2020Altschul}*	\\	[6pt]	
	$		(k_\pi)_{XX}-(k_\pi)_{YY},(k_\pi)_{(XY)}		$&$	< 10^{-23}	$&		Chiral perturbation theory		&	\rr{2016Kamand}*	\\	[2pt]	
	$		(k_\pi)_{(XZ)},(k_\pi)_{(YZ)}		$&$	< 10^{-24}	$&	\dit{	Chiral perturbation theory	}	&	\rr{2016Kamand}*	\\	[2pt]	
	$		\de^\pi		$&$	> -7 \times 10^{-13}	$&		Astrophysics		&	\rr{2016Altschul}*	\\	[2pt]	
	$	\dit{	\de^\pi	}	$&$	(-1.5 \mbox{ to } 200) \times 10^{-11}	$&	\dit{	Astrophysics	}	&	\rr{2007BAPion}*	\\	[2pt]	
	$		|c^\pi|		$&$	< 10^{-10}	$&	\dit{	Astrophysics	}	&	\rr{hepph0610324}*	\\
\end{tabular}}												
			\end{center}												

			\begin{center}\scalebox{0.95}{												
			\begin{tabular}{@{}cr@{\hspace{12pt}}ll@{}}												
			\multicolumn{4}{c}{Table \DaQuarkDimFiveUp.\ Quark sector, $d\geq 5$ (part 1 of 2)}\\												
			\hlineone												
			{\bf Combination}		&	\multicolumn{1}{c}{{\bf Result}}	&		\multicolumn{1}{c}{{\bf System}}		&	\multicolumn{1}{c}{{\bf Ref.}}	\\		
			\hlinetwo												
	$		\ring{m}^u_0,\ \ring{m}^u_2		$&$	< 3 \times 10^{-18} \gev^{-1}	$&		Astrophysics		&	\rr{2025Petrov}*	\\		
	$		\ring{a}^u_0,\ \ring{a }^u_2		$&$	< 2 \times 10^{-29} \gev^{-1}	$&	\dit{	Astrophysics	}	&	\rr{2025Petrov}*	\\		
	$		\ring{m}^d_0,\ \ring{m}^d_2		$&$	< 6 \times 10^{-18} \gev^{-1}	$&	\dit{	Astrophysics	}	&	\rr{2025Petrov}*	\\		
	$		\ring{a}^d_0,\ \ring{a }^d_2		$&$	< 9 \times 10^{-29} \gev^{-1}	$&	\dit{	Astrophysics	}	&	\rr{2025Petrov}*	\\		
	$		a_{\text{S}u}^{(5)TTT}		$&$	(-4.1 \mbox{ to } 12) \times 10^{-8}\gev^{-1}	$&		Drell-Yan process		&	\rr{2024lunghi}*	\\	[2pt]	
	$		a_{\text{S}u}^{(5)TZZ} 		$&$	(-22 \mbox{ to } 7.1) \times 10^{-8}\gev^{-1}	$&	\dit{	Drell-Yan process	}	&	\rr{2024lunghi}*	\\
\end{tabular}}												
			\end{center}												
															
			\begin{center}\scalebox{0.95}{												
			\begin{tabular}{@{}cr@{\hspace{12pt}}ll@{}}												
			\multicolumn{4}{c}{Table \DaQuarkDimFiveUp.\ Quark sector, $d\geq 5$ (part 2 of 2)}\\												
			\hlineone												
			{\bf Combination}		&	\multicolumn{1}{c}{{\bf Result}}	&		\multicolumn{1}{c}{{\bf System}}		&	\multicolumn{1}{c}{{\bf Ref.}}	\\		
			\hlinetwo												
	$		|\kfk_{TTT} - 0.97\kfk_{TTZ}|		$&$	< 2 \times 10^{-24}\gev^{-1}	$&		$K$ oscillations		&	\rr{2019ek}*	\\	[2pt]	
	$		|\kfk_{TTX}|,~|\kfk_{TTY}|		$&$	< 1.8 \times 10^{-25}\gev^{-1}	$&	\dit{	$K$ oscillations	}	&	\rr{2019ek}*	\\	[2pt]	
	$		\kfk_{TTT}		$&$	(-2.4 \pm 4.4) \times 10^{-17}\gev^{-1}	$&	\dit{	$K$ oscillations	}	&	\rr{2019ek}*	\\	[2pt]	
	$		\kfk_{TTX}		$&$	(1.2 \pm 2.8) \times 10^{-18}\gev^{-1}	$&	\dit{	$K$ oscillations	}	&	\rr{2019ek}*	\\	[2pt]	
	$		\kfk_{TTY}		$&$	(2.7 \pm 2.7) \times 10^{-18}\gev^{-1}	$&	\dit{	$K$ oscillations	}	&	\rr{2019ek}*	\\	[2pt]	
	$		\kfk_{TTZ}		$&$	(4.2 \pm 3.0) \times 10^{-18}\gev^{-1}	$&	\dit{	$K$ oscillations	}	&	\rr{2019ek}*	\\
\end{tabular}}												
			\end{center}												

		\begin{center}\scalebox{0.95}{										
		\begin{tabular}{@{}cr@{\hspace{12pt}}ll@{}}										
		\multicolumn{4}{c}{Table \DaElweak.\ Electroweak sector, $d=3,4$}\\										
		\hlineone										
		{\bf Combination}	&	\multicolumn{1}{c}{{\bf Result}}	&		\multicolumn{1}{c}{{\bf System}}		&	\multicolumn{1}{c}{{\bf Ref.}}	\\	
		\hlinetwo										
	$	(k_1)_\mu, (k_2)_\mu	$&$	< 10^{-4} \gev	$&		$Z$ boson decay		&	\rr{2025Williams}*	\\	[2pt]
	$	|(2 k_1)^\mu (2k_1)_\mu)|	$&$	< 7.2\times  10^{-21} m_e^2	$&		Lepton magnetic moments		&	\rr{2023Hurtado-Silva}*	\\	[2pt]
	$	|(2 k_2)^\mu (2k_2)_\mu)|	$&$	< 3.2\times  10^{-19} m_e^2	$&	\dit{	Lepton magnetic moments	}	&	\rr{2023Hurtado-Silva}*	\\	[2pt]
	$	|(\Re k_\ph)^\mu (\Re k_\ph)_\mu)|	$&$	< 3.29\times  10^{-29} \gev^2	$&	\dit{	Lepton magnetic moments	}	&	\rr{2022Castulo}*	\\	[2pt]
	$	|(k_{\ph\ph})^{01}|	$&$	< 10^{-12}	$&		Laser interferometry		&	\rr{2018Hernandez}*	\\	[2pt]
	$	|(k_{\ph\ph})^{02}|	$&$	< 10^{-12}	$&	\dit{	Laser interferometry	}	&	\rr{2018Hernandez}*	\\
\end{tabular}}										
		\end{center}										

		\begin{center}\scalebox{0.95}{											
		\begin{tabular}{@{}cr@{\hspace{12pt}}ll@{}}											
		\multicolumn{4}{c}{Table \DaGluon.\ Gluon sector, $d=4$}\\											
		\hlineone											
		{\bf Combination}	&	\multicolumn{1}{c}{{\bf Result}}	&		\multicolumn{1}{c}{{\bf System}}		&	\multicolumn{1}{c}{{\bf Ref.}}	\\		
		\hlinetwo											
	$	|k^{TT}|	$&$	\thy{< 10^{-21}}	$&		Various		&	\rr{2017Noordmans}*	\\		
	$	|k^{TJ}|	$&$	\thy{< 10^{-19}}	$&	\dit{	Various	}	&	\rr{2017Noordmans}*	\\		
	$	|k^{JK}|	$&$	\thy{< 10^{-27}}	$&	\dit{	Various	}	&	\rr{2017Noordmans}*	\\		
	$	|k^{XX}|, |k^{YY}|	$&$	\thy{< 10^{-27}}	$&	\dit{	Various	}	&	\rr{2017Noordmans}*	\\		
	$	|k^{ZZ}|	$&$	\thy{< 10^{-20}}	$&	\dit{	Various	}	&	\rr{2017Noordmans}*	\\		
	$	|k^a| \mbox{ for } a=1, 2, 8, 9, 10	$&$	\thy{< 10^{-34}}	$&		Astrophysics		&	\rr{2017Noordmans}*	\\		
	$	|k^a| \mbox{ for } a=3, 4, 5, 6, 7	$&$	\thy{< 10^{-35}}	$&	\dit{	Astrophysics	}	&	\rr{2017Noordmans}*	\\		
	$	|\kt_{\rm tr}^{\rm QCD}|	$&$	\thy{< 2 \times 10^{-13}}	$&	\dit{	Astrophysics	}	&	\rr{2006Carone}*	\\		
		\hlinethree											
		\end{tabular}}											
		\end{center}											

			\begin{center}\scalebox{0.95}{															
			\begin{tabular}{@{}cr@{\hspace{12pt}}ll@{}}															
			\multicolumn{4}{c}{Table \DaGravDimTwo.\ Gravity sector, $d=2$}\\															
			\hlineone															
			{\bf Combination}		&		\multicolumn{1}{c}{{\bf Result}}		&		\multicolumn{1}{c}{{\bf System}}		&		\multicolumn{1}{c}{{\bf Ref.}}		\\	
			\hlinetwo															
	$		\kjm 2 I {00}		$&$		(-0.5 \mbox{ to } 1.3) \times 10^{-63} \gev^2		$&		Gravitational waves		&		\rr{2025Wang}*		\\	
			\hlinethree															
			\end{tabular}}															
			\end{center}															
																		
			\begin{center}\scalebox{0.95}{															
			\begin{tabular}{@{}cr@{\hspace{12pt}}ll@{}}															
			\multicolumn{4}{c}{Table \DaGravDimThree.\ Gravity sector, $d=3$ (part 1 of 2)}\\															
			\hlineone															
			{\bf Combination}		&		\multicolumn{1}{c}{{\bf Result}}		&		\multicolumn{1}{c}{{\bf System}}		&		\multicolumn{1}{c}{{\bf Ref.}}		\\	
			\hlinetwo															
	$		\al(\overline{a}_{\rm eff}^{e+p})_T		$&$		(-1.3 \pm 8.5)\times 10^{-5} \gev		$&		Gravitational redshift		&		\rr{2025Qin}*		\\	
	$		\dit{\al(\overline{a}_{\rm eff}^{e+p})_T}		$&$		(-0.18 \pm 2.32)\times 10^{-5} \gev		$&	\dit{	Gravitational redshift	}	&		\rr{2025Qin}*		\\	
	$		\al(\overline{a}_{\rm eff}^{n})_T		$&$		(-1.3 \pm 8.5)\times 10^{-5} \gev		$&	\dit{	Gravitational redshift	}	&		\rr{2025Qin}*		\\	
	$		\dit{\al(\overline{a}_{\rm eff}^{n})_T}		$&$		(-0.18 \pm 2.32)\times 10^{-5} \gev		$&	\dit{	Gravitational redshift	}	&		\rr{2025Qin}*		\\	
	$		\al(\overline{a}_{\rm eff}^{e+p})_T		$&$		(-3 \pm 9)\times 10^{-5} \gev		$&		Null redshift		&		\rr{2025Qin}*		\\	
	$		\al(\overline{a}_{\rm eff}^{n})_T		$&$		(-3 \pm 8)\times 10^{-5} \gev		$&	\dit{	Null redshift	}	&		\rr{2025Qin}*		\\
\end{tabular}}															
			\end{center}															
																		
			\begin{center}\scalebox{0.95}{															
			\begin{tabular}{@{}cr@{\hspace{12pt}}ll@{}}															
			\multicolumn{4}{c}{Table \DaGravDimThree.\ Gravity sector, $d=3$ (part 2 of 2)}\\															
			\hlineone															
			{\bf Combination}		&		\multicolumn{1}{c}{{\bf Result}}		&		\multicolumn{1}{c}{{\bf System}}		&		\multicolumn{1}{c}{{\bf Ref.}}		\\	
			\hlinetwo															
	$		|\al(\overline{a}_{\rm eff}^{e+p})_T|		$&$		<10^{-7} \gev		$&		Free-fall WEP		&		\rr{2010akjt}*		\\	
	$		\dit{|\al(\overline{a}_{\rm eff}^{e+p})_T|}		$&$		<10^{-10} \gev		$&		Force-comparison WEP		&		\rr{2010akjt}*		\\	
	$		|\al(\overline{a}_{\rm eff}^n)_T|		$&$		<10^{-7} \gev		$&		Free-fall WEP		&		\rr{2010akjt}*		\\	
	$		\dit{|\al(\overline{a}_{\rm eff}^n)_T|}		$&$		<10^{-10} \gev		$&		Force-comparison WEP		&		\rr{2010akjt}*		\\	
	$		|\al(\overline{a}_{\rm eff}^{e+p-n})_T-\frac 1 3 m^p (\overline{c}^{e+p-n})_{TT}|		$&$		<10^{-8} \gev		$&		Combined WEP		&		\rr{2010akjt}*		\\	
	$		|\al(\overline{a}_{\rm eff}^{e+p-n})_T-\frac 1 3 m^p (\overline{c}^{e+p-n})_{TT}+(\frac 1 2 \cos^2 \ch - \frac 1 6)m^n(\overline{c}^n)_Q|		$&$		<10^{-8} \gev		$&		Free-fall WEP		&		\rr{2010akjt}*		\\
\end{tabular}}															
			\end{center}															

			\begin{center}\scalebox{0.95}{															
			\begin{tabular}{@{}cr@{\hspace{12pt}}ll@{}}															
			\multicolumn{4}{c}{Table \DaGravDimFour.\ Gravity sector, $d=4$ (part 1 of 5)}\\															
			\hlineone															
			{\bf Combination}		&		\multicolumn{1}{c}{{\bf Result}}		&		\multicolumn{1}{c}{{\bf System}}		&		\multicolumn{1}{c}{{\bf Ref.}}		\\	
			\hlinetwo															
	$		 \sol{ 4 }{ 0 0 } 		$&$		-0.07^{+1.67}_{-0.97}		$&		Gravitational waves		&		\rr{2023Ray}*		\\	
	$		 \sol{ 4 }{ 1 0 } 		$&$		-0.28^{+1.80}_{-1.96}		$&	\dit{	Gravitational waves	}	&		\rr{2023Ray}*		\\	
	$		 \Re \sol{ 4 }{ 1 1 } 		$&$		-0.18^{+0.56}_{-0.32}		$&	\dit{	Gravitational waves	}	&		\rr{2023Ray}*		\\	
	$		 \Im \sol{ 4 }{ 1 1 } 		$&$		0.52^{+0.69}_{-0.53}		$&	\dit{	Gravitational waves	}	&		\rr{2023Ray}*		\\	
	$		 \sol{ 4 }{ 2 0 } 		$&$		-0.44^{+1.95}_{-1.46}		$&	\dit{	Gravitational waves	}	&		\rr{2023Ray}*		\\	
	$		 \Re \sol{ 4 }{ 2 1 } 		$&$		0.36^{+0.84}_{-0.45}		$&	\dit{	Gravitational waves	}	&		\rr{2023Ray}*		\\
\end{tabular}}															
			\end{center}															
																		
			\begin{center}\scalebox{0.95}{															
			\begin{tabular}{@{}cr@{\hspace{12pt}}ll@{}}															
			\multicolumn{4}{c}{Table \DaGravDimFour.\ Gravity sector, $d=4$ (part 2 of 5)}\\															
			\hlineone															
			{\bf Combination}		&		\multicolumn{1}{c}{{\bf Result}}		&		\multicolumn{1}{c}{{\bf System}}		&		\multicolumn{1}{c}{{\bf Ref.}}		\\	
			\hlinetwo															
	$		\sb^{TX}		$&$		(3.4 \pm 7.9)\times 10^{-6}		$&		Atom Interferometry		&		\rr{2023Zhang}		\\	
	$		\sb^{TY}		$&$		(7.8 \pm 12.1)\times 10^{-6}		$&	\dit{	Atom Interferometry	}	&		\rr{2023Zhang}		\\	
	$		\sb^{TZ}		$&$		(-7.2 \pm 9.8)\times 10^{-6}		$&	\dit{	Atom Interferometry	}	&		\rr{2023Zhang}		\\	
	$		\sb^{XX}-\sb^{YY}		$&$		(1.8 \pm 2.1)\times 10^{-9}		$&	\dit{	Atom Interferometry	}	&		\rr{2023Zhang}		\\	
	$		\sb^{XY}		$&$		(1.1 \pm 1.0)\times 10^{-9}		$&	\dit{	Atom Interferometry	}	&		\rr{2023Zhang}		\\	
	$		\sb^{XZ}		$&$		(5.9 \pm 5.5)\times 10^{-10}		$&	\dit{	Atom Interferometry	}	&		\rr{2023Zhang}		\\
\end{tabular}%
}										
	\end{center}										

\end{document}